\documentclass[12pt]{article}
\usepackage{layout,amssymb,amsmath,epsfig}

\begin{document}
\title{\bf  Study of Anisotropic Compact Stars in Starobinsky Model}
\author{M. Zubair$^1$ \thanks{mzubairkk@gmail.com; drmzubair@ciitlahore.edu.pk} and G.
Abbas$^2$ \thanks{abbasg91@yahoo.com}\\\\
$^1$ Department of Mathematics, COMSATS\\
Institute of Information Technology, Lahore, Pakistan.\\
$^2$ Department of Mathematics, The Islamia  \\
University of Bahawalpur, Bahawalpur, Pakistan.}

\date{}

\maketitle
\begin{abstract}
The aim of this paper is to study the formation of anisotropic
compact stars in modified $f(R)$ theory of gravity, which is the
generalization of the Einstein's gravity. To this end, we have used
the solution of Krori and Barua to the anisotropic distribution of
matter in $f(R)$ gravity. Further, we have matched the interior
solution with the exterior solution to determine the constants of
Krori and Barua solution. Finally the constant have been determined
by using the data of compact compact stars like 4$U1820-30, Her X-1,
SAX J 1808-3658$. Using the evaluated form of the solutions, we have
discussed the regularity of matter components at the center as well
as on the boundary, energy conditions, anisotropy, stability
analysis and mass-radius relation of the compact stars 4$U1820-30$,
$Her X-1$, $SAX J 1808-3658.$
\end{abstract}

\section{Introduction}

In the weak field regime, General Relativity (GR) has succeeded to
counter the observation tests whereas strong filed is yet to be
explored. In fact the great success of GR has not stopped the
alternatives being proposed and modifications begin to appear in
very early days of this theory. Current investigations reveals the
fact GR fails to explain the strong gravitational field effects
which suggests that this theory may require modification. The
presence of higher order term in Einstein-Hilbert (EH) action has
motivated the researcher to modify this theory in strong field
regime. In 1980, Starobinsky presented the idea of curvature driven
inflationary scenario, where the action of GR is replaced by
$f(R)=R+\lambda{R}^2$ \cite{1}. In current years, numerous efforts
have been made to ge beyond the original Einstein theory, in order
to discuss the accelerated expansion of the universe in more
scientific ways \cite{2}.

Exploring the exotic compact objects in modified gravity, would be a
scientific tool to handle this problem. The study of strong
gravitational field of compact objects clearly explain the
significant differences between GR and its modification. The
modeling of massive star in $f(R)$ gravity have added some
additional proprieties to stars \cite{3,4}. According to Psaltis
\cite{5} the strong gravitational fields could be considered as
modified theories of gravity if we consider GR as the weak field
limit of some more realistic effective gravitational theory. One can
consider the stability of relativistic stars in $f(R)$ gravity as a
test of the theory's viability; some $f(R)$ models do not allow the
existence of stable star configurations and thus are considered
unrealistic \cite{6}. However, possible problems regarding the
existence of these objects may be avoided due to the so-called
Chameleon Mechanism \cite{7}. The study of neutron stars in $f(R)$
gravity is currently an active field and people have worked on their
existence as well as the stability \cite{8}-\cite{8d}.

During the last decades many researchers have derived the models of
anisotropic compact stars. Egeland \cite{9} discussed the modeling
the mass-radius relation of of the Neutron star and concluded that a
cosmological constant would exist due to density of the vacuum. For
this purpose, Egeland used the equation of hydrostatic equilibrium
and fermion gas equation of state (EoS). Using spherical symmetry of
compact stars, an exact solution of equation of was proposed by Mak
and Harko \cite{15}, which predicts the properties of strange stars.
Rahaman et al. \cite{17} provided the extension of Krori-Barua
\cite{18} models using the Chaplygin gas EOS. Lobo \cite{19}
investigated the models of the compact objects with a barotropic
EOS. He also extended the Mazur-Mottola gravastar models by using
the junction conditions between static spacetime and Schwarzschild
vacuum solution. In the present study, we have investigated the
formation of spherically symmetric anisotropic compact stars in
$f(R)$ gravity that were initially suggested by Alcok et al.
\cite{20} and Haensel et al.\cite{21}. The anisotropic compact stars
models with linear equation of state and cvariable cosmological
constant have been formulated by Hossein et al. \cite{28a}.

We study the formation of anisotropic compact stars with more
generalized $f(R)$ model i.e., $f(R)=R+\lambda{R}^2$ (where
$\lambda$ is constant) and conclude that $f(R)$ gravity can provide
that existence the of anisotropic compact stars candidates X-ray
bruster 4$U1820-30, X-ray$ pulsar $Her X-1,$ Millisecond pulsar $SAX
J 1808-3658$. The objective of this paper is that if compact star
solutions exist in $f(R)$, what are the constraints on $f(R)$ model
and parameters of the theory? The spherically symmetric models of
the compact stars proposed here are associated with $f(R)$ theory of
gravity and we analyzed the stability of these models by using the
anisotropic property of the model.  This paper is organized as
follow. In the coming section, we formulate the equations of motion
for anisotropic source and static metric in $f(R)$ gravity. In
Section \textbf{3}, we discuss the implementation of the solution to
a class of compact stars and present the physical behavior of the
proposed models. In the last section, we summarize the findings of
the paper.

\section{Anisotropic Matter Configuration in $f(R)$ Gravity}

The action of $f(R)$ theory of gravity in the presence of matter is
given by \cite{5}
\begin{equation}\label{1}
\mathcal{\mathcal{I}}=\int{dx^4\sqrt{-g}[f(R)+\mathcal{L}_{(matter)}]},
\end{equation}
where $8{\pi}G=1,~R$ is the scalar curvature, $f(R)$ is an arbitrary
function of $R$ as well as its higher powers and
$\mathcal{L}_{(matter)}$ denotes the Lagrangian density of matter
part. Hence, we get the following form of field equations
\begin{equation}\label{2}
G_{{\mu}{\nu}}=R_{{\mu}{\nu}}-\frac{1}{2}Rg_{{\mu}{\nu}}
=T_{{\mu}{\nu}}^{(curv)}+T_{{\mu}{\nu}}^{(matter)},
\end{equation}
where $T_{{\mu}{\nu}}^{(matter)}$ is the stress-energy tensor of the
matter and $T_{{\mu}{\nu}}^{(curv)}$ is curvature term , given by
\begin{equation}\label{3}
T_{{\mu}{\nu}}^{(curv)}=\frac{1}{F(R)}\left[\frac{1}{2}g_{\mu\nu}(f(R)-RF(R))
+F(R)^{;{\alpha}{\beta}}(g_{\mu\alpha}g_{\nu\beta}-g_{\mu\nu}g_{\alpha\beta})\right],
\end{equation}
where $F(R)=f^{'}(R)$.

The general spherically symmetric metric is given by
\begin{equation}\label{4}
ds^2=-e^{\mu(r)}dt^2+e^{\nu(r)}dr^2+r^2(d\theta^2+sin^2\theta{d\phi^2}),
\end{equation}
where $\nu=Ar^2$, $\mu=Br^2+C$  \cite{18}, $A$, $B$ and $C$ are
constants.

For the anisotropic fluid the energy-momentum tensor is defined by
\begin{equation}\label{5}
T^m_{\alpha\beta}=(\rho+p_t)u_\alpha{u}_\beta-p_tg_{\alpha\beta}+(p_r-p_t)v_\alpha{v}_\beta,
\end{equation}
where $u_\alpha=e^\frac{\mu}{2}\delta^0_\alpha$,
$v_\alpha=e^\frac{\nu}{2}\delta^0_\alpha$, are four velocities,
$\rho$ is energy density,  $p_r$ and $p_t$ are radial and transverse
pressures, respectively. In this case set of field equations is
\begin{eqnarray}\nonumber
\rho&=&-e^{-\nu}F''+e^{-\nu}\left(\frac{\nu'}{2}-\frac{2}{r}\right)F'
+\frac{e^{-\nu}}{r^2}\left(\frac{\mu''r^2}{2}+\frac{\mu'^2r^2}{4}
-\frac{\mu'\nu'r^2}{4}+\mu'r\right)F\\\label{6}&-&\frac{1}{2}f,\\\label{7}
p_r&=&
e^{-\nu}\left(\frac{\mu'}{2}+\frac{2}{r}\right)F'-\frac{e^{-\nu}}{r^2}
\left(\frac{\mu''r^2}{2}+\frac{\mu'^2r^2}{4}-\frac{\mu'\nu'r^2}{4}
-\nu'r\right)F+\frac{1}{2}f,\\\nonumber
p_t&=&-e^{-\nu}F''+e^{-\nu}\left(\frac{\mu'}{2}-\frac{\nu'}{2}+\frac{1}{r}\right)F'
-\frac{e^{-\nu}}{r^2}\left(\frac{\mu'r}{2}-\frac{\nu'r}{2}-e^{\nu}+1\right)F\\\label{8}&+&\frac{1}{2}f.
\end{eqnarray}

The Starobinsky model is \cite{1}
\begin{equation}\label{9}
f(R)=R+\lambda{R}^2,
\end{equation}
where $\lambda$ is an arbitrary constant. The most important thing
in existence of compact stars is the requirement of static
configuration i.e., the EoS satisfies the condition $\rho-3p>0$.
Therefor, in fixing $\lambda$, one needs to analyze this situation
and avoid the existence of singularities. In this settings we find
that the viable values of $\lambda$ lies in the range $0<\lambda<6$.
One can choose suitable value of $\lambda$ according to this
condition. Herein we set $\lambda=2km^2$.

For this model the equation (\ref{6})-(\ref{8}) become
\begin{eqnarray}\nonumber
\rho&=&\frac{e^{-2\nu}}{8r^4}\{r^4\lambda\mu'^4-2r^3\lambda\mu'^3
(-4+r\nu')+r^2\lambda\mu'^2(16+8r\nu'-11r^2\nu'^2+4r^2\mu''\\\nonumber&+&8r^2\nu'')+4r^2\lambda
\mu'(-16r\nu'^2+3r^2\nu'^3+\nu'(-4+9r^2\mu''-7r^2\nu'')+2r(-2\mu''\\\nonumber&+&6\nu''
-2r\mu'''+r\nu'''))+4(-2e^\nu{r}^2+2e^{2\nu}r^2-20\lambda+24e^\nu
\lambda-4e^{2\nu}\lambda\\\nonumber&+&12r^3\lambda\nu'^3-3r^4\lambda\mu''^2-r^2\lambda
\nu'^2(12+11r^2\mu'')+8r^2\lambda\nu''+8r^4\lambda\mu''\mu''-16\\\nonumber&\times&r^3\lambda\mu'''+
2r\nu'(e^\nu{r}^2-8\lambda+16r^2\lambda\mu''-14r^2\lambda\nu''+6r^3\lambda\mu''')+
8r^3\lambda\nu'''\\\label{10}&-&4r^4\lambda\mu^{(iv)})\},\\\nonumber
p_r&=&\frac{e^{-2\nu}}{8r^4}\{-r^4\lambda\mu'^4-2r^4\lambda\mu'^3\nu'+r^3
\lambda\mu'^2(-24\nu'+3r\nu'^2+4r(\mu''-\nu''))\\\nonumber&-&4(-2e^\nu{r}^2+2e^{2\nu}r^2+28\lambda
-24e^\nu\lambda-4e^{2\nu}\lambda-12r^2\lambda\nu'^2-16r^2\lambda\mu''\\\nonumber&+&
8r^3\lambda\nu'\mu''+r^4\lambda\mu''^2+16r^2\lambda\nu''-8r^3
\lambda\mu'''+8r\mu'(e^\nu{r}^2-8\lambda+3r^2\lambda\nu'^2\\\label{11}&+&6r^2
\lambda\mu''-r\lambda\nu'(8+r^2\mu'')-4r^2\lambda\nu''+r^3\lambda\mu''')\},\\\nonumber
p_t&=&\frac{e^{-\nu}}{8r^4}\{r^4\lambda\mu'^4+2r^3\lambda\mu'^3(2-3r\nu')+
r^2\mu'^2(-32r\lambda\nu'+17r^2\lambda\nu'^2+2(e^\nu{r}^2\\\nonumber&-&8\lambda+6r^2
\lambda\mu''-6r^2\lambda\nu''))-2r\mu'(-38r^2\lambda\nu'^2+6r^3
\lambda\nu'^3+r\nu'(e^\nu{r}^2-24\lambda\\\nonumber&+&28r^2\lambda\mu''-14r^2
\lambda\nu'')-2(e^\nu{r}^2-4\lambda+12e^nu\lambda+10r^2\lambda\mu''-
14r^2\lambda\nu''\\\nonumber&+&6r^3\lambda\mu'''-2r^3\lambda\nu'''))-4(12r^3
\lambda\nu'^3-11r^4\lambda\mu''\nu'^2-5r^4\lambda\mu''^2+\mu''(-e^\nu
r^4\\\nonumber&+&8r^2\lambda+8r^4\lambda\nu'')+r\nu(e^\nu{r}^2-28\lambda+12e^\nu
\lambda+28r^2\lambda\mu''-28r^2\lambda\nu''+12r^3\\\label{12}&\times&\lambda\mu''')+4\lambda
(-7+6e^\nu+e^{2\nu}-3r^3\mu'''+2r^3\nu'''-r^4\mu^{(iv)}))\}.
\end{eqnarray}
We have five unknown functions $\rho, p_r, p_t, \mu, \nu$, and three
Eqs.(\ref{10})-(\ref{12}). We have to chose any two functions,
keeping in mind the regularity conditions of the compact stars, we
chose $\mu$ and $\nu$ as after Eq.(\ref{4}). As metric functions are
exponential i.e., $e^{\mu(r)}, e^{\nu(r)}$,  for  $\nu=Ar^2$,
$\mu=Br^2+C$, metric functions remain exponential as well as regular
even at center of the star. Moreover, this choice satisfies the
boundary conditions in the center of the star \cite{8d}
\begin{equation}\nonumber
\rho(0)=\rho_c,\quad \nu(0)=0, \quad \frac{d\mu}{dr}(0)=0.
\end{equation}

From the metric potential function, we get following form of matter
components
\begin{eqnarray}\nonumber
\rho&=&\frac{1}{r^4}e^{-2Ar^2}\{e^{2Ar^2}(r^2-2\lambda)+2(-5-3B^2r^4+6B^3r^6
+B^4r^8+12A^3r^6\\\nonumber&\times&(2+Br^2)-A^2r^4(40+68Br^2+11B^2r^4)+A(-4r^2
+48Br^4\\\label{13}&+&26B^2r^6
-2B^3r^8))\lambda+e^{Ar^2}(-r^2+2Ar^4+12\lambda)\},\\\nonumber
p_r&=&\frac{1}{r^4}e^{-2Ar^2}\{-e^(2Ar^2)(r^2-2\lambda)+2(-7+11B^2r^4+
2B^3r^6-B^4r^8\\\nonumber&+&3A^2r^4(2+Br^2)^2-2Ar^2(4+16Br^2+9B^2r^4+
B^3r^6))\lambda+e^{Ar^2}(r^2\\\label{14}&+&2Br^4+12\lambda)\},\\\nonumber
p_t&=&\frac{1}{r^4}e^{-2Ar^2}\{-2(-7+6e^{Ar^2}+e^{2Ar^2})\lambda+16B^3r^6\lambda
+2B^4r^8\lambda-24A^3r^6(2\\\nonumber&+&Br^2)\lambda+2A^2r^4(28+74Br^2+17B^2r^4)\lambda+
B^2r^4(e^{Ar^2}r^2+22\lambda)+2Br^2\\\nonumber&\times&(-6\lambda+e^{Ar^2}(r^2+6\lambda))-
Ar^2(4(-7+19Br^2+25B^2r^4+3B^3r^6)\lambda\\\label{15}&+&e^{Ar^2}(r^2+Br^4+12\lambda))\}.
\end{eqnarray}
Here, we consider the following form of linear equation of state
(EOS)
\begin{equation}
p_r=w_r\rho,\quad\quad\quad\quad \quad p_t=w_t \rho
\end{equation}

The above equations lead to the following relations
\begin{eqnarray}\nonumber
\omega_r&=&\{-e^{2Ar^2}(r^2-2\lambda)+2(-7+11B^2r^4+2B^3r^6-B^4r^8+3A^2r^4(2+Br^2)^2
\\\nonumber&-&2Ar^2(4+16Br^2+9B^2r^4+B^3r^6))\lambda+e^{A
r^2}(r^2+2Br^4+12\lambda)\}/\{e^{2Ar^2}\\\nonumber&\times&(r^2-2\lambda)+2(-5-3B^2r^4+
6B^3r^6+B^4r^8+12A^3r^6(2+Br^2)-A^2r^4\\\nonumber&\times&(40+68Br^2+11B^2r^4)+A(-4r^2
+48Br^4+26B^2r^6-2B^3r^8))\lambda\\\label{16}&+&e^{Ar^2}(-r^2+2Ar^4+12\lambda)\},\\\nonumber
\omega_t&=&\{-2(-7+6e^{Ar^2}+e^{2Ar^2})\lambda+16B^3r^6\lambda+2B^4r^8\lambda
-24A^3r^6(2+Br^2)\lambda\\\nonumber&+&2A^2r^4(28+74Br^2+17B^2r^4)\lambda+B^2r^4(e^{Ar^2}r^2+22\lambda)
+2Br^2(-6\lambda\\\nonumber&+&e^{Ar^2}(r^2+6\lambda))-Ar^2(4(-7+19Br^2+25B^2r^4+3B^3r^6)\lambda+e^{Ar^2}
(r^2\\\nonumber&+&Br^4+12\lambda))\}/\{e^{2Ar^2}(r^2-2\lambda)+2(-5-3B^2r^4+6B^3r^6+B^4r^8
\\\nonumber&+&12A^3r^6(2+Br^2)-A^2r^4(40+68Br^2+11B^2r^4)+A(-4r^2+48Br^4\\\label{17}&+&26B^2r^6-2B^3r^8))\lambda
+e^{Ar^2}(-r^2+2Ar^4+12\lambda)\}.
\end{eqnarray}

\section{Physical Analysis}
In this section, we discuss the following physical properties of the
solutions.

\subsection{Anisotropic Constraints}
\begin{figure}
\centering \epsfig{file=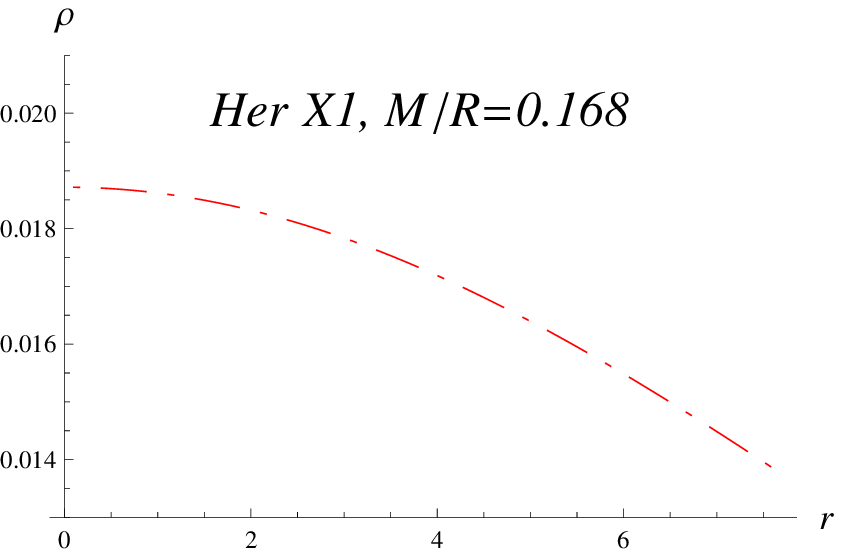, width=.34\linewidth,
height=1.4in}\epsfig{file=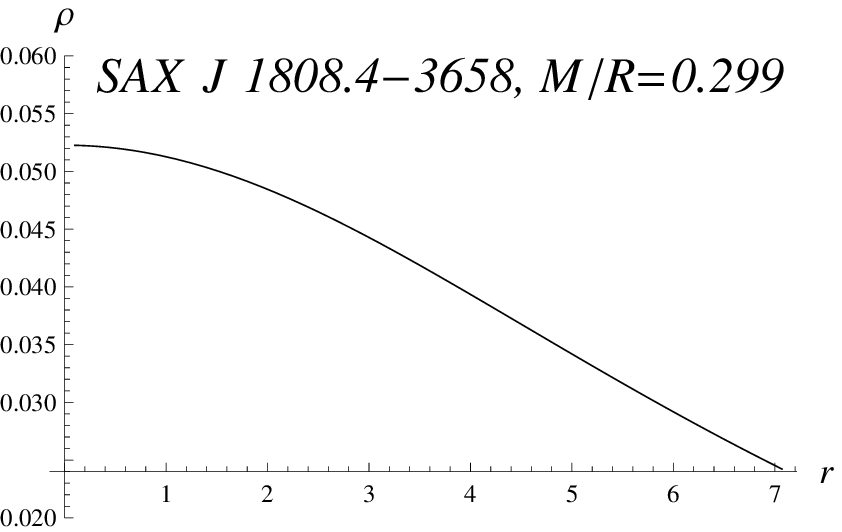, width=.36\linewidth,
height=1.4in}\epsfig{file=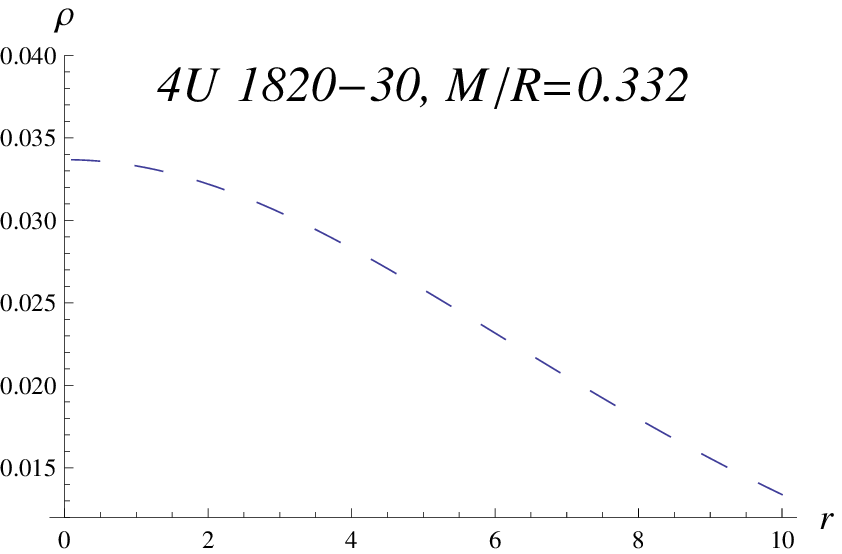, width=.34\linewidth,
height=1.4in}\caption{Variation of energy density $\rho$ versus
radial coordinate $r(km)$. Herein, we set $\lambda=2km^2$.}
\end{figure}
\vspace*{2ex}
\begin{figure}
\centering \epsfig{file=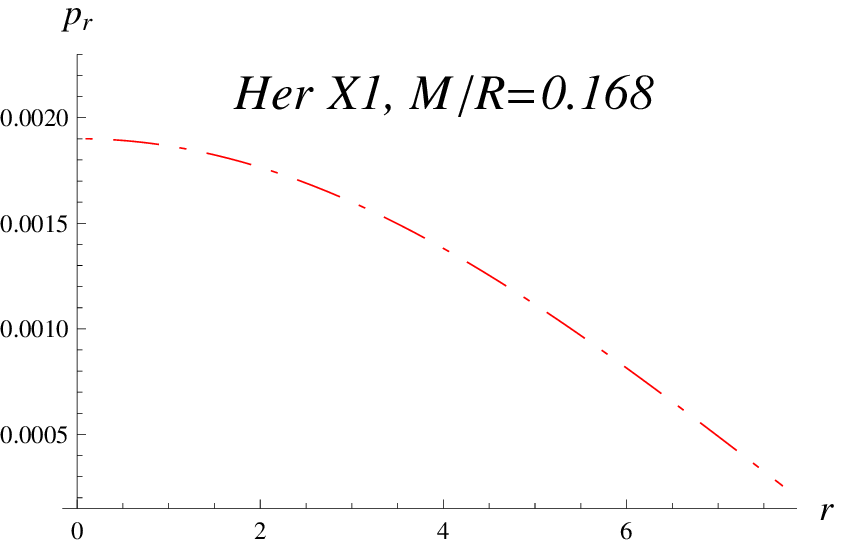, width=.34\linewidth,
height=1.4in}\epsfig{file=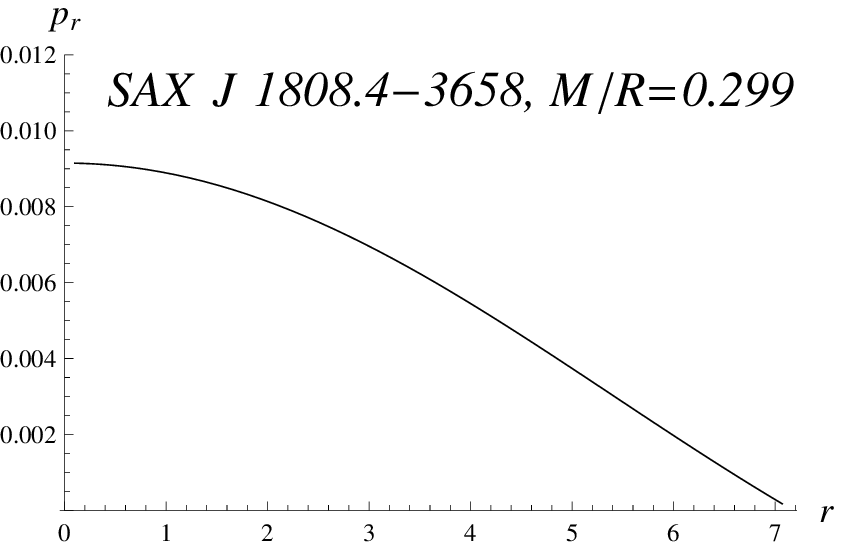, width=.36\linewidth,
height=1.4in}\epsfig{file=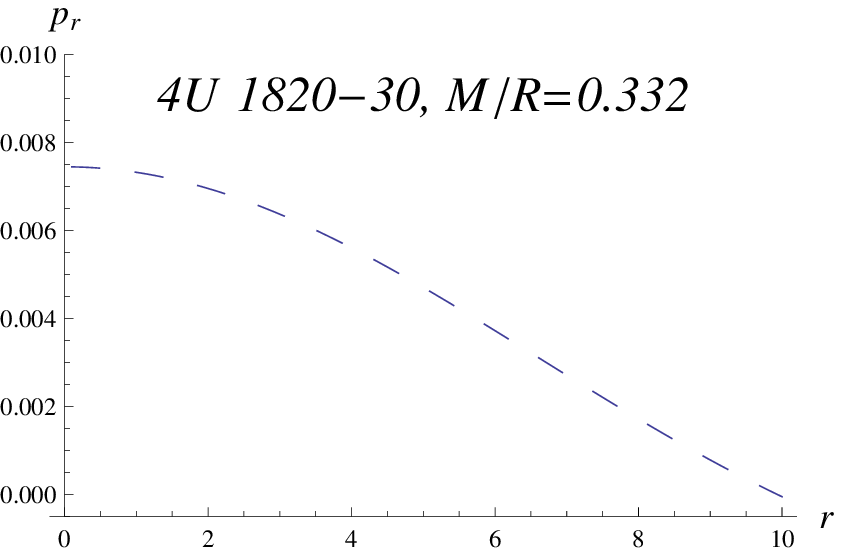, width=.34\linewidth,
height=1.4in}\caption{Variation of radial pressure $p_r$ versus
radial coordinate $r(km)$ .}
\end{figure}
\begin{figure}
\centering \epsfig{file=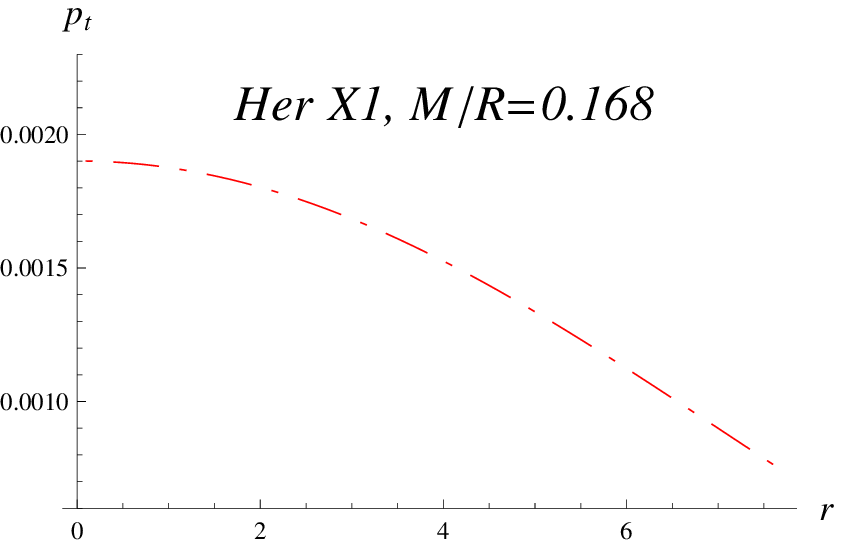, width=.34\linewidth,
height=1.4in}\epsfig{file=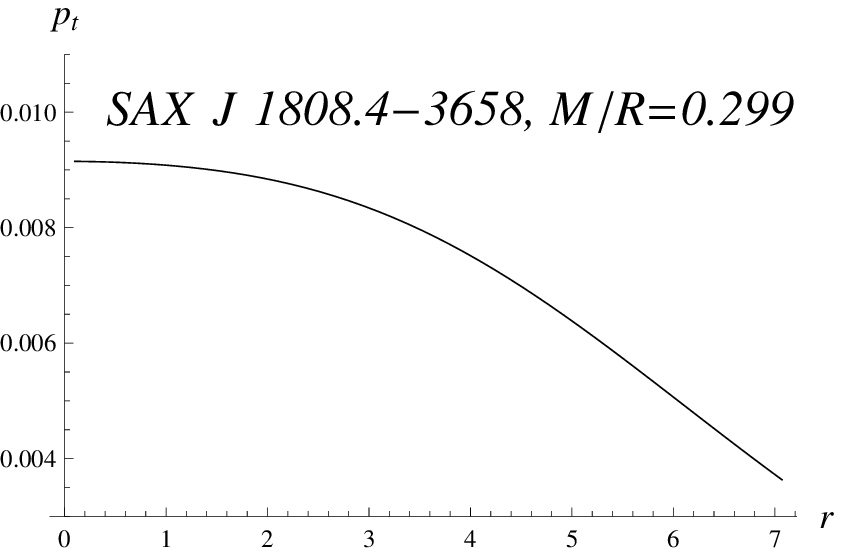, width=.36\linewidth,
height=1.4in}\epsfig{file=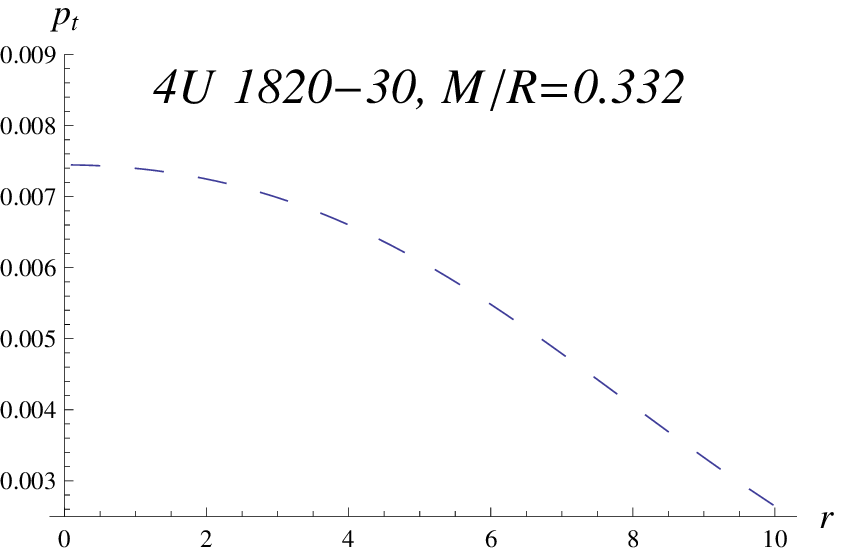, width=.34\linewidth,
height=1.4in}\caption{Variation of transverse pressure $p_t$ versus
radial coordinate $r(km)$ .}
\end{figure}

In the first place, we present the evolution of energy density
$\rho$, radial pressure $p_r$ and tangential pressure $p_t$ as shown
in Figures \textbf{1-3} for different strange stars (see Table 1).

Taking derivatives of equations (\ref{13}) and (\ref{14}) with
respect to radial coordinate, we have
\begin{eqnarray}\nonumber
\frac{d\rho}{dr}&=&\frac{1}{r^5}2e^{-2Ar^2}\{-e^{2Ar^2}(r^2-4\lambda)
-4(-5-3B^3r^6-B^4r^8+12A^4r^8(2\\\nonumber&+&Br^2)-A^3r^6(52+80Br^2+11B^2r^4)+A^2r^4(-4
+82Br^2+37B^2r^4\\\nonumber&-&2B^3r^6)+Ar^2(-7-16B^2r^4+8B^3r^6+ B^4
r^8))\lambda+e^{Ar^2}(Ar^4-2A^2r^6\\\label{20}&-&24\lambda+r^2(1-12A\lambda))\},\\\nonumber
\frac{dp_r}{dr}&=&\frac{1}{r^5}2e^{-2Ar^2}\{e^{2Ar^2}(r^2-4\lambda)-4(-7-B^3r^6
+B^4r^8+3A^3r^6(2+Br^2)^2\\\nonumber&-&A^2r^4(8+38Br^2+21B^2r^4+2B^3r^6)+Ar^2(-11
+20B^2r^4+4B^3r^6\\\label{21}&-&B^4r^8))\lambda-e^{Ar^2}(Ar^4+2ABr^6+24\lambda+r^2(1+12A\lambda))\}.
\end{eqnarray}
The evolution of $\frac{d\rho}{dr}$ and $\frac{dp_r}{dr}$ is shown
in Figures \textbf{4} and \textbf{5}. It can be seen that
$\frac{d\rho}{dr}<0$ and $\frac{dp_r}{dr}<0$.
\begin{figure}
\centering \epsfig{file=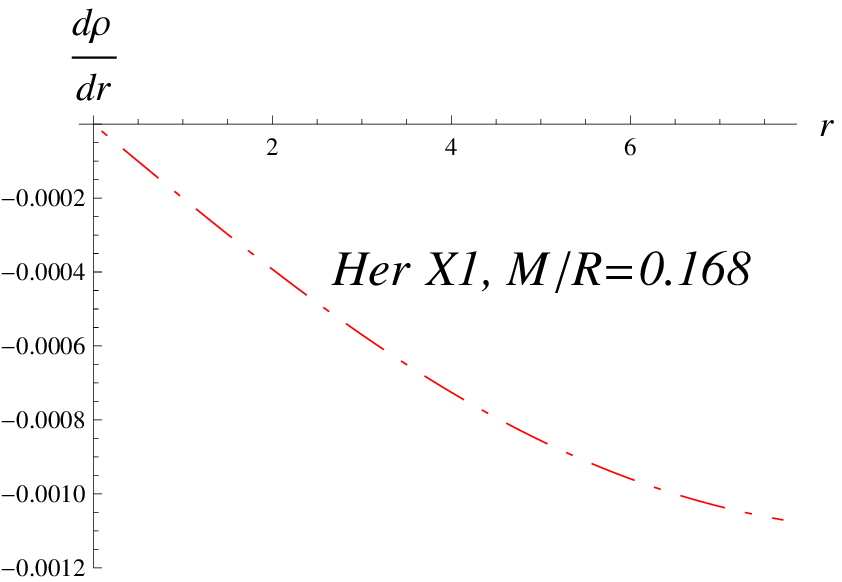, width=.34\linewidth,
height=1.4in}\epsfig{file=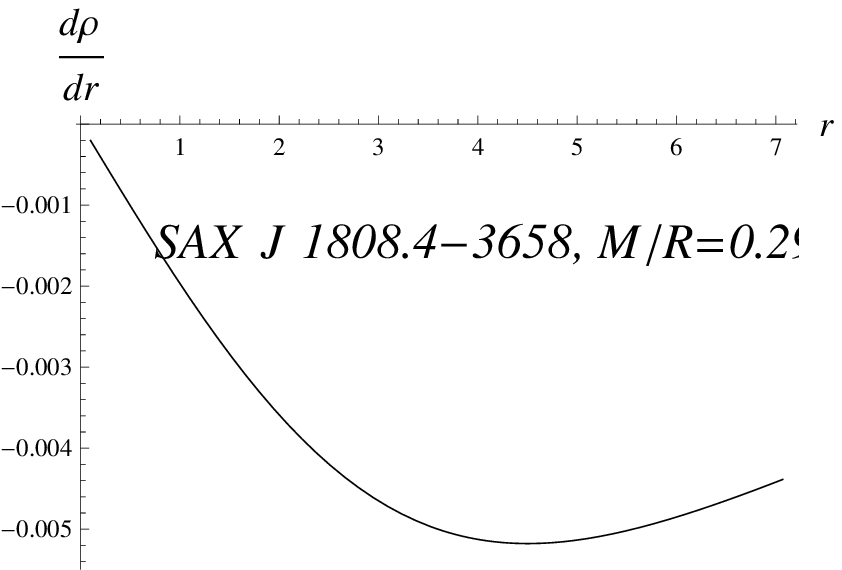, width=.36\linewidth,
height=1.4in}\epsfig{file=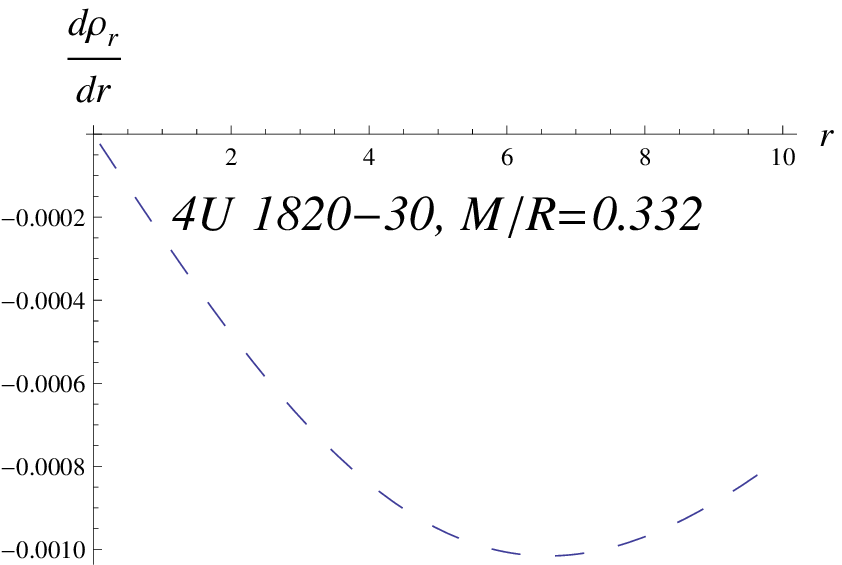, width=.34\linewidth,
height=1.4in}\caption{Behavior of $\frac{d\rho}{dr}$ versus radial
coordinate $r(km)$.}
\end{figure}
\begin{figure}
\centering \epsfig{file=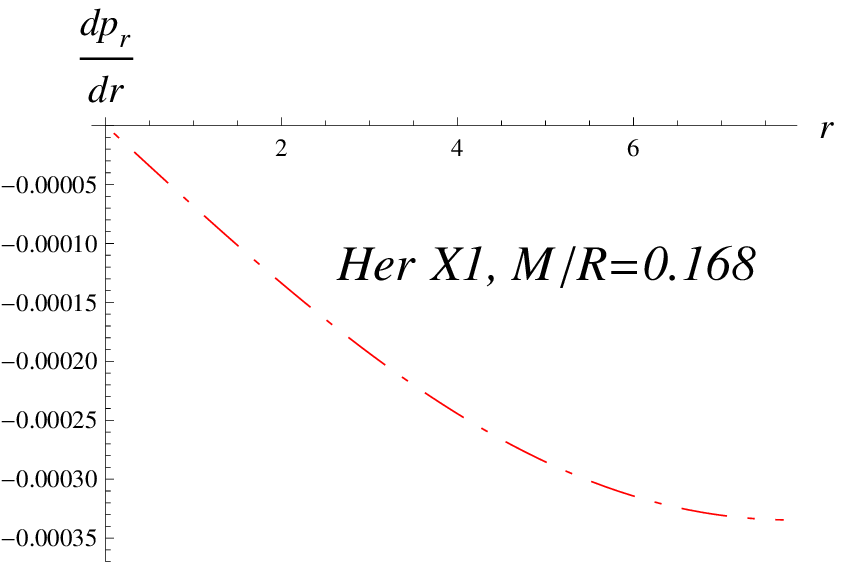, width=.34\linewidth,
height=1.4in}\epsfig{file=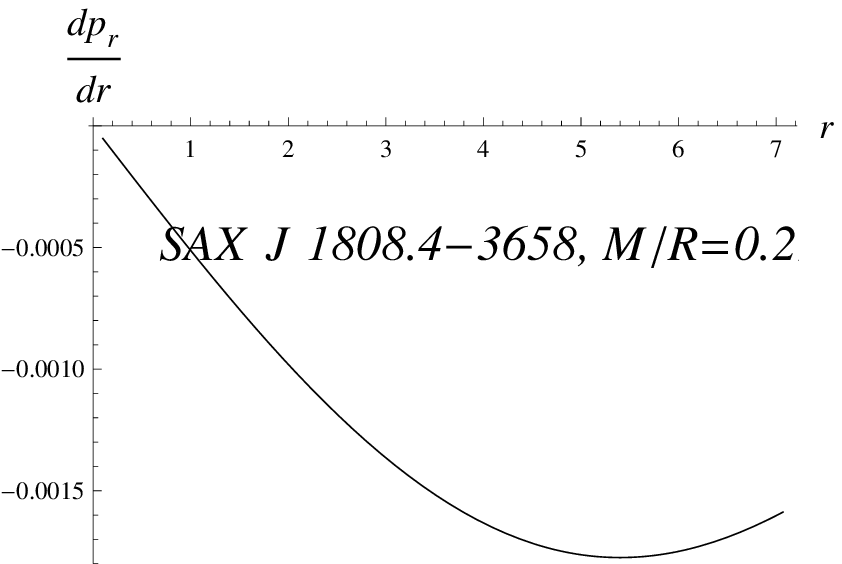, width=.36\linewidth,
height=1.4in}\epsfig{file=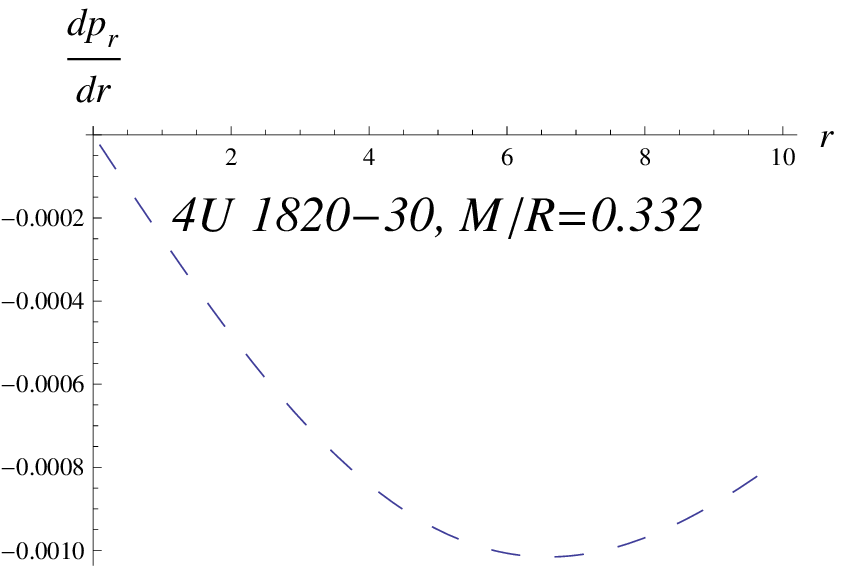, width=.34\linewidth,
height=1.4in}\caption{Behavior of $\frac{p_r}{dr}$ versus radial
coordinate $r(km)$ .}
\end{figure}

We also examine the behavior of derivatives of $\rho$ and $p_r$ at
center $r=0$ of compact star and it is found that
\begin{eqnarray}\nonumber
\frac{d\rho}{dr}=0, \quad \quad \frac{dp_r}{dr}=0,\\\label{zz}
\frac{d^2\rho}{dr^2}<0, \quad \quad \frac{d^2p_r}{dr^2}<0.
\end{eqnarray}
Equation (\ref{zz}) shows the maximality of central $\rho$ and
$p_r$. Hence $\rho$ and $p_r$ attain maximum values at $r=0$ and
functional values decreases with the increase in $r$ as shown in
Figures \textbf{1-3}. We present the evolution of EoS parameters
$\omega_r$ and $\omega_t$ in Figures \textbf{6} and \textbf{7} for
different strange stars. We call these parameters as effective since
these involve the contribution from the additional terms in $f(R)$
gravity. Here, it is clear that, like normal matter distribution,
the bound on the effective EOS in this case is given by
$0<\omega_i(r)<1$, ($i=r,t$).
\begin{figure}
\centering \epsfig{file=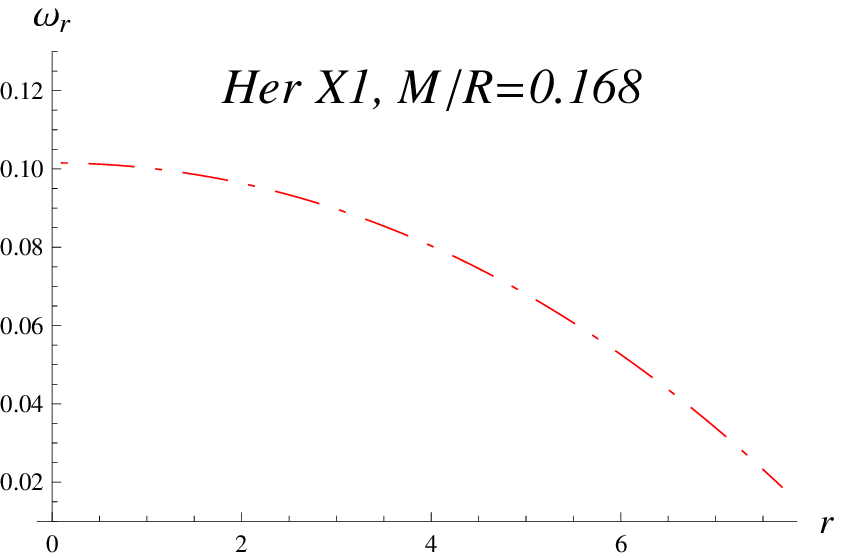, width=.34\linewidth,
height=1.4in}\epsfig{file=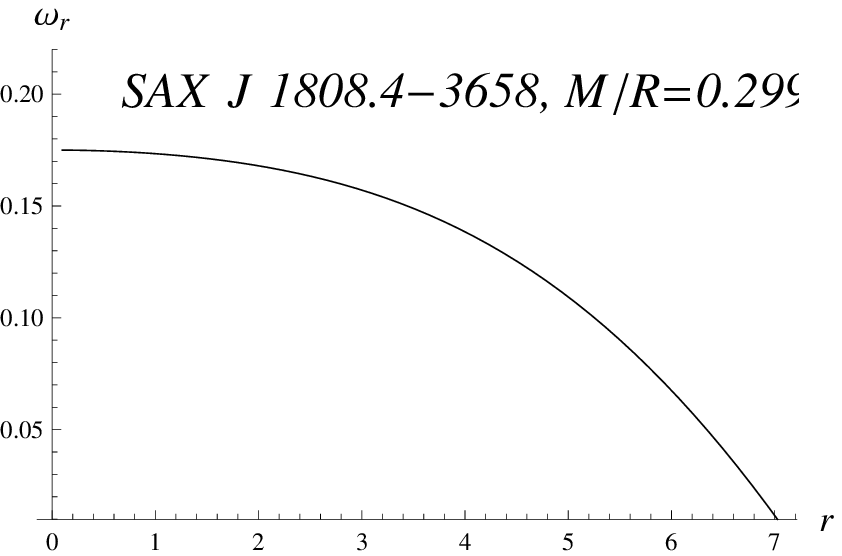, width=.36\linewidth,
height=1.4in}\epsfig{file=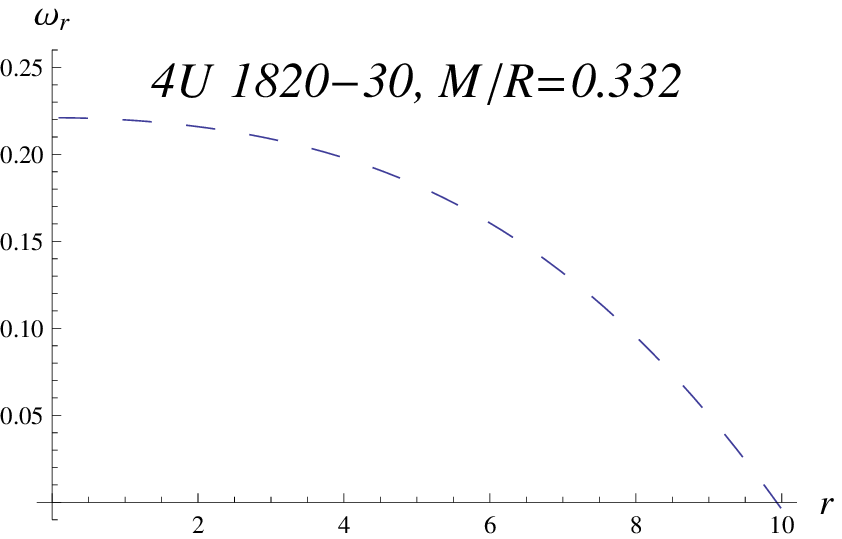, width=.34\linewidth,
height=1.4in}\caption{Variation of EoS parameter $\omega_r$ versus
radial coordinate $r(km)$ .}
\end{figure}
\begin{figure}
\centering \epsfig{file=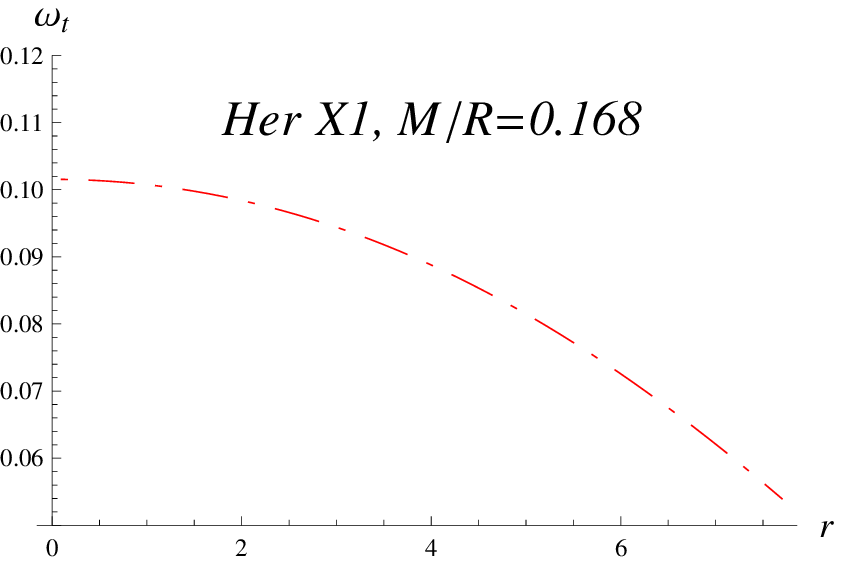, width=.34\linewidth,
height=1.4in}\epsfig{file=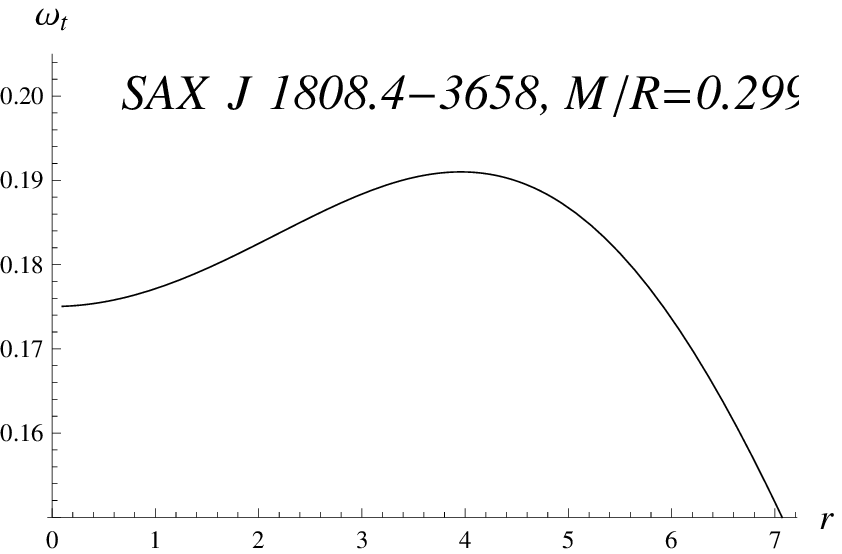, width=.36\linewidth,
height=1.4in}\epsfig{file=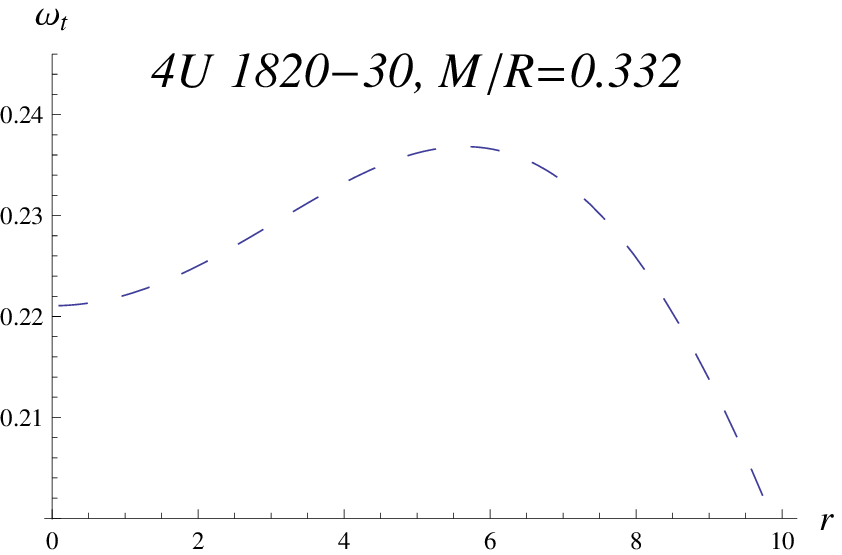, width=.34\linewidth,
height=1.4in}\caption{Variation of EoS parameter $\omega_t$ versus
radial coordinate $r(km)$ .}
\end{figure}

The anisotropy measurement $\Delta=\frac{2}{r}(p_t-p_r)$ for this
model is given by
\begin{eqnarray}\nonumber
\Delta&=&\frac{1}{r^5}2e^{-2Ar^2}\{e^{2Ar^2}(r^2-4\lambda)-4(-7+3Br^2-3B^3r^6
-B^4r^8+6A^3r^6(2\\\nonumber&+&Br^2)-A^2r^4(8+31Br^2+7B^2r^4)+Ar^2(-11+3Br^2+16B^2r^4
+2B^3\\\label{23}&\times&r^6))\lambda-e^{Ar^2}(Ar^4+(A-B)Br^6+24\lambda+r^2(1
+12A\lambda-12B\lambda))\}.
\end{eqnarray}
The measure of anisotropy is directed outward when $p_t>p_r$ which
implies $\Delta>0$ whereas it is directed inward if $p_t<p_r$
resulting in $\Delta<0$. In this discussion we consider the
fractional pressure anisotropy given by $\Delta{r}/p_r$. The
evolution of fractional pressure anisotropy is shown in Figure 8. It
is obvious that $\Delta{r}/p_r$ remains positive at the stellar
interior of strange star candidates. Hence for this case repulsive
force exists which allows the construction of more massive
configuration.
\begin{figure}
\centering \epsfig{file=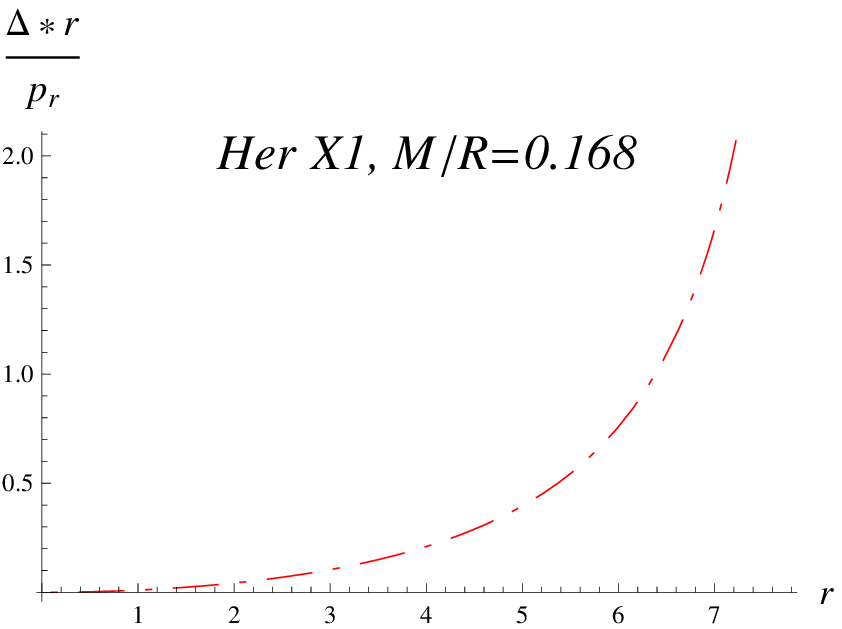, width=.34\linewidth,
height=1.4in}\epsfig{file=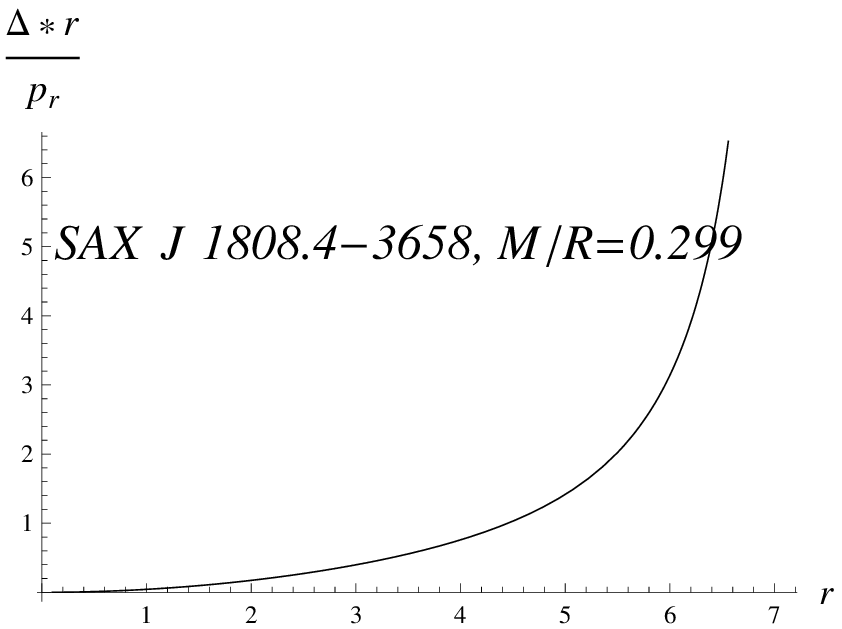, width=.36\linewidth,
height=1.4in}\epsfig{file=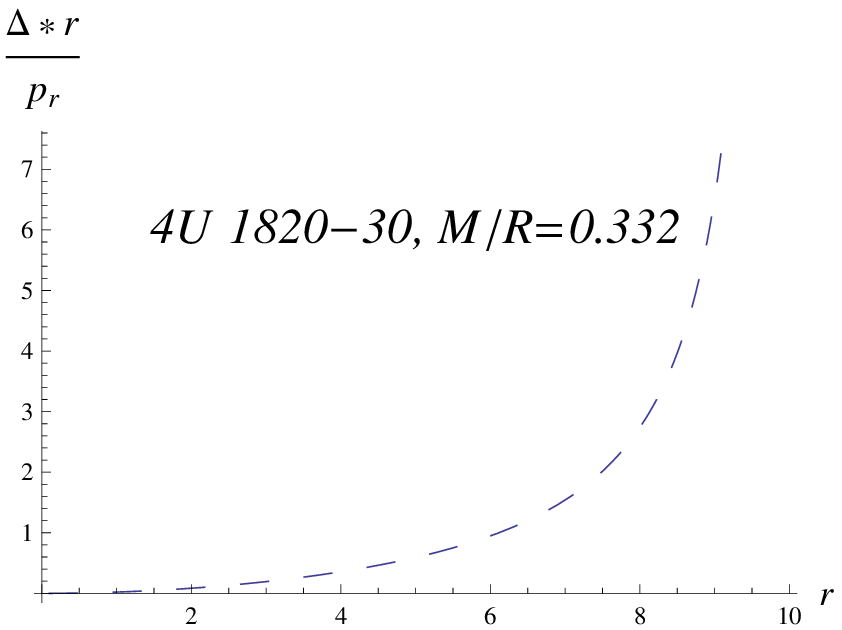, width=.34\linewidth,
height=1.4in}\caption{Variation of anisotropy measurement .}
\end{figure}

It is interesting to see that anisotropy vanishes at the center
$r=0$ and the corresponding pressures take the form
$p_t(0)=p_r(0)=p_0=34A^2\lambda+2B(1+11B\lambda)-A(1+64B\lambda)$.

\subsection{Matching Conditions}
 In \cite{29b}, Cooney et al. studied the formation of compact
 objects like Neutron Star in
$f(R)$ gravity theories with perturbation constraints. The
Schwartzchild-de Sitter metric is considered as an exterior solution
which is matched with the interior spherical symmetry using
conditions analogous to that in GR. According to these authors
\cite{29a}-\cite{29aa} Schwarzschild solution is the most suitable
solution as exterior geometry of the star. Using this approach, a
lot of work has been done \cite{8,8a,8b,8c,8d, 29a}-\cite{29e} by
takind Schwarzschild or Vaidya metric to address the problems
related to gravitational collapse and neutron stars in $f(R)$
gravity.

The vacuum exterior spherically symmetric metric given by
\begin{equation}\label{21}
 ds^2=-\left(1-\frac{2M}{r}\right)dt^2+\left(1-
 \frac{2M}{r}\right)^{-1}dr^2+r^2d\theta^2+r^2 sin^2{\theta}d\varphi^2,
\end{equation}
The continuity of the metric functions $g_{tt}$, $g_{rr}$ and
$\frac{\partial g_{tt}}{\partial r}$ yield,
\begin{eqnarray}\label{22}
  g_{tt}^-=g_{tt}^+,~~~~~
   g_{rr}^-=g_{rr}^+,~~~~~
   \frac{\partial g_{tt}^-}{\partial r}=\frac{\partial g_{tt}^+}{\partial r},
  \end{eqnarray}
where $-$ and $+$, are quantities for for the internal and external
portion of star. Hence we get
 \begin{eqnarray}\label{23}
  A&=&-\frac{1}{R^2}ln\left(1-\frac{2M}{R}\right),\\\label{24}
 B&=&\frac{M}{R^3}{{\left(1-\frac{2M}{R}\right)}^{-1}},\\\label{24a}
 C&=&ln\left(1-\frac{2M}{R}\right)-\frac{M}{R}{{\left(1-\frac{2M}{R}\right)}^{-1}}.
\end{eqnarray}
Li et al. \cite{1*} studied X-ray pulsar SAX J1808.4-3658 to compare
its mass-radius relation with theoretical mass-radius relation of
strange star and for neutron star candidates and shown the
consistency of strange star model with SAX J1808.4-3658. They
suggested that SAX J1808.4-3658 is a likely strange star candidate
and calculated masses and radii of strange star as $1.44M_\odot$,
$1.32M_\odot$ and $7.07km$, $6.53km$, respectively. Zhang et al.
\cite{2*} presented the mass measurement for the neutron star in 4U
1820-30 and reported mass of the order $\simeq2.2M\odot$. In
\cite{3*}, mass and radius of neutron star in 4U 1820-30 are
determined with $1\sigma$ error as $M =1.58\pm0.06M_\odot$ and a
radius of $R=9.11\pm0.4km$. However, upper bound limit in this
measurement is consistent with that in \cite{2*}. In fact there is a
certain uncertainty in measurement of mass and radius of a compact
stars. Abubekerov et al. \cite{1*} estimated the mass of Her X-1
using more recent and physically justified techniques and found two
different values of masses $m_x=0.85\pm0.15M_\odot$ and
$m_x=1.8M_\odot$ through the radial-velocity curves. This
uncertainty may be due to the tense X-ray heating in Her X-1. For
$M$ and $R$ \cite{1*}-\cite{29g} of the compact stars, the constants
$A$ and $B$ are given in the table \textbf{1}.
\begin{table}[ht]
\caption{Values of constants for given Masses and Radii of Stars}
\begin{center}
\begin{tabular}{|c|c|c|c|c|c|}
\hline {Strange Quark Star}&  \textbf{ $M$} & \textbf{$R(km)$} &
\textbf{ $\frac{M}{R}$} &\textbf{ $A(km ^{-2})$}& \textbf{$B(km
^{-2})$}
\\\hline  $Her X-1$& 0.88$M_\odot$& 7.7&0.168&0.0069062764281 &
$0.0042673646183$
\\\hline $SAX J 1808.4-3658$& 1.435$M_\odot$& 7.07&0.299& 0.018231569740 &
$0.014880115692$
\\\hline $4U 1820-30$&2.25$M_\odot$& 10.0 &0.332&0.010906441192 &
$0.0098809523811$
\\\hline
\end{tabular}
\end{center}
\end{table}

\subsection{Energy Conditions}
The validity of these energy conditions is necessary for a
physically reasonable energy-momentum tensor. The energy conditions
for anisotropic fluid are defined by the following relations
\begin{eqnarray}
{NEC}:\quad&&\rho+p_r\geq0, \quad \rho+p_t\geq0,\\
{WEC}:\quad&&\rho\geq0, \quad \rho+p_r\geq0, \quad \rho+p_t\geq0,\\
{SEC}:\quad&&\rho+p_r\geq0, \quad
\rho+p_t\geq0, \quad \rho+p_r+2p_t\geq0,\\
{DEC}:\quad&&\rho>|p_r|, \quad \rho>|p_t|.
\end{eqnarray}
In Figure \textbf{9} energy conditions are fulfilled for our model.
\begin{figure}
\centering \epsfig{file=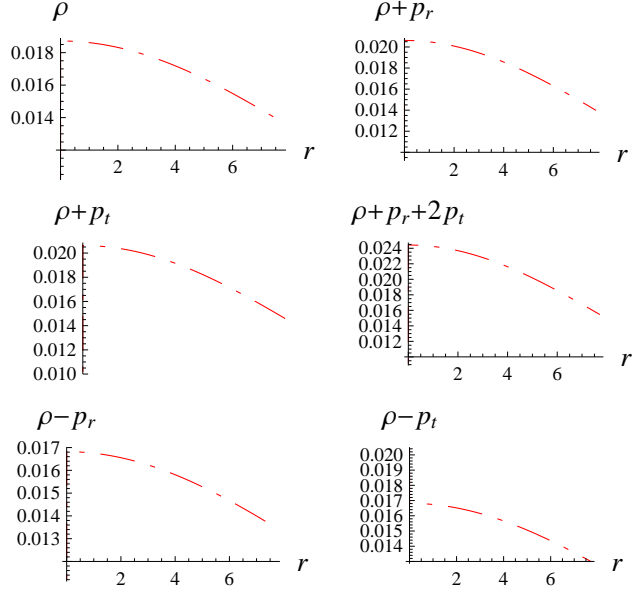}\caption{Evolution of energy
constraints for compact star Her X-1.}
\end{figure}

\subsection{TOV Equation}

The generalized Tolman-Oppenheimer-Volkoff (TOV) equation gets the
form
\begin{equation}\label{27}
\frac{dp_r}{dr}+\frac{\nu'(\rho+p_r)}{2}+\frac{2(p_r-p_t)}{r}=0
\end{equation}
Following \cite{28a}, above equation can be written as
\begin{eqnarray}\nonumber
&&F_g+F_h+F_a=0, \\\label{28} && F_g=-Br(\rho+p_r),\quad
F_h=-\frac{dp_r}{dr}, \quad F_a=\frac{2(p_t-p_r)}{r}
\end{eqnarray}
Using the effective $\rho$, $p_r$ and $p_t$ (\ref{10})-(\ref{12}),
for strange star Her X-1, we have plotted these values in figure
\textbf{10}.

\begin{figure}
\centering \epsfig{file=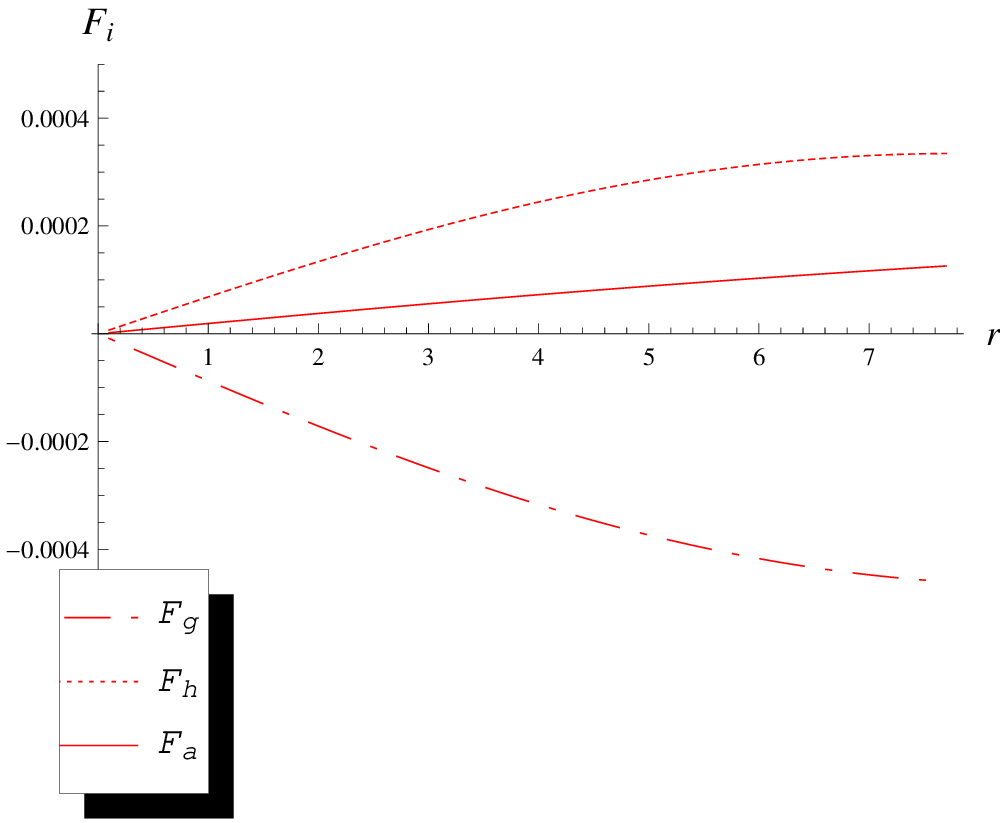, width=.34\linewidth,
height=1.4in}\epsfig{file=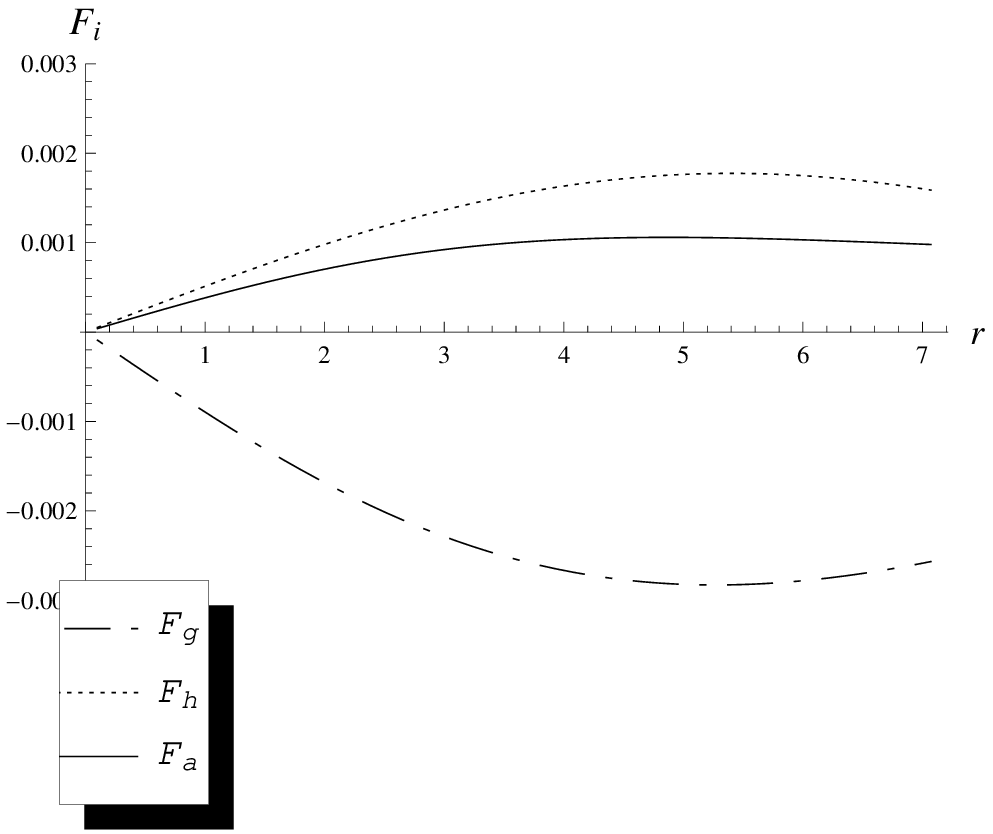, width=.36\linewidth,
height=1.4in}\epsfig{file=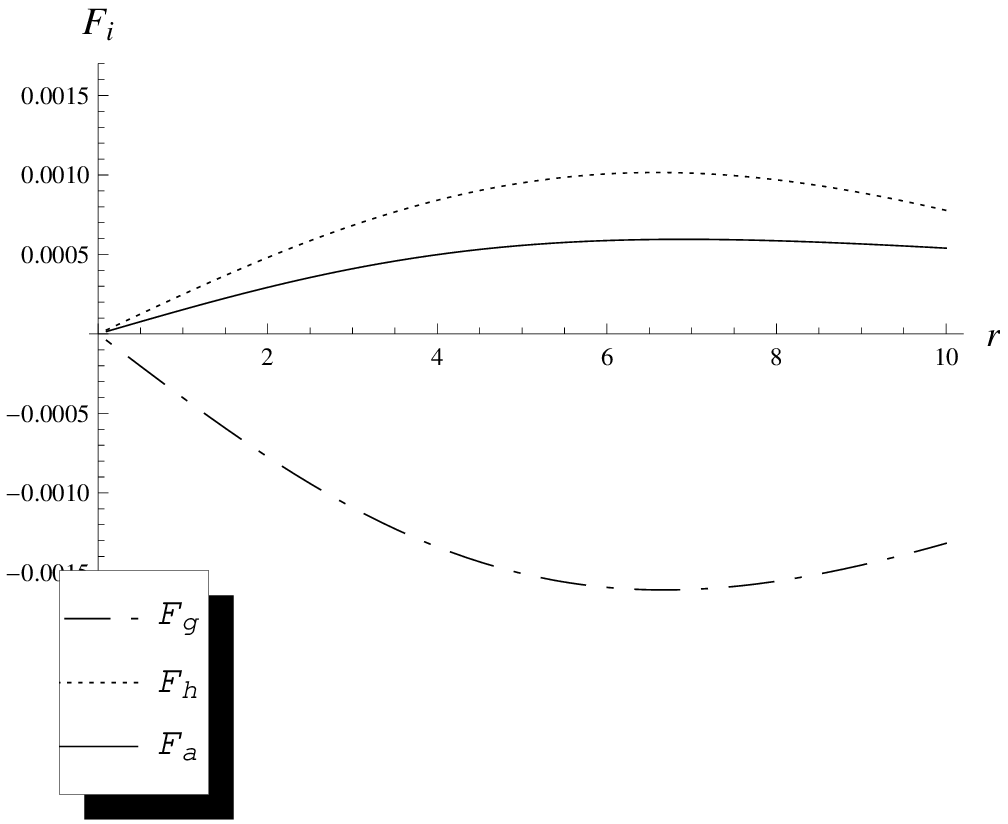, width=.34\linewidth,
height=1.4in}\caption{Variation of gravitating, hydrostatic and
pressure anisotropic forces for compact star candidates.}
\end{figure}

\subsection{Stability Analysis}

People \cite{30}-\cite{32} have discussed the appearance of cracking
in spherical compact objects by using different approaches. Herrera
\cite{30} introduced the concept of cracking to identify potentially
unstable anisotropic matter configuration. It was considered to
explain the behavior of fluid distribution, once the equilibrium
configuration has been perturbed and total non-vanishing radial
forces of different signs appear within the system. Now, by
considering the sound speeds one can assess the potentially stable
and unstable regions established through the difference of sound
propagation within the matter configuration. The region for which
radial sound of sound $v^2_{sr}$ is greater than the transverse
speed of sound $v^2_{st}$ is potentially stable.

To analyze the stability of our model we calculate the radial and
transverse speeds as
\begin{eqnarray}\nonumber
v^2_{sr}&=&\{-e^{2Ar^2}(r^2-4\lambda)+4(-7-B^3r^6+B^4r^8+3A^3r^6(2+Br^2)^2
-A^2r^4\\\nonumber&\times&(8+38Br^2+21B^2r^4+2B^3r^6)+Ar^2(-11+20B^2r^4+4
B^3r^6\\\nonumber&-&B^4r^8))\lambda+e^{Ar^2}(Ar^4+2ABr^6+24\lambda+r^2(1+12A\lambda))\}
/\{e^{2Ar^2}(r^2\\\nonumber&-&4\lambda)+4(-5-3B^3r^6-B^4r^8+12A^4r^8(2+Br^2)
-A^3r^6(52+80Br^2\\\nonumber&+&11B^2r^4)+A^2r^4(-4+82Br^2+37B^2r^4-2B^3r^6)
+Ar^2(-7-16B^2r^4\\\label{24}&+&8B^3r^6+B^4r^8))\lambda+e^{A
r^2}(-Ar^4+2A^2r^6+24\lambda+r^2(-1+12A\lambda))\},\\\nonumber
v^2_{st}&=&\{B^2e^{Ar^2}r^6+4(-7+6e^{Ar^2}+e^{2Ar^2})\lambda-12B(-1+e^{Ar^2})r^2\lambda
+16B^3r^6\lambda\\\nonumber&+&4B^4r^8\lambda+48A^4r^8(2+Br^2)\lambda-4A^3r^6(40+86Br^2
+17B^2r^4)\lambda+A^2r^4\\\nonumber&\times&(4(-14+75Br^2+67B^2r^4+6B^3r^6)\lambda+e^{Ar^2}(r^2
+Br^4+12\lambda))-Ar^2\\\nonumber&\times&(-8(-7+3e^{Ar^2})\lambda+56B^3r^6\lambda+4B^4r^8\lambda
+B^2r^4(e^{Ar^2}r^2+144\lambda)+3Br^2\\\nonumber&\times&(-8\lambda+e^{Ar^2}(r^2+4\lambda)))\}
/\{-e^{2Ar^2}(r^2-4\lambda)-4(-5-3B^3r^6-B^4r^8\\\nonumber&+&12A^4r^8(2+Br^2)-A^3r^6(52
+80Br^2+11B^2r^4)+A^2r^4(-4+82Br^2\\\nonumber&+&37B^2r^4-2B^3r^6)+Ar^2(-7-16B^2r^4+8B^3r^6
+B^4r^8))\lambda+e^{Ar^2}(Ar^4\\\label{25}&-&2A^2r^6-24\lambda+r^2(1-12A\lambda))\}.
\end{eqnarray}
In Figures \textbf{11} and \textbf{12} it is shown that $v^2_{sr}$
and $v^2_{st}$ satisfy the inequalities $0\leq{v}^2_{sr}\leq1$ and
$0\leq{v}^2_{st}\leq1$ within the anisotropic matter configuration.

\begin{figure}
\centering \epsfig{file=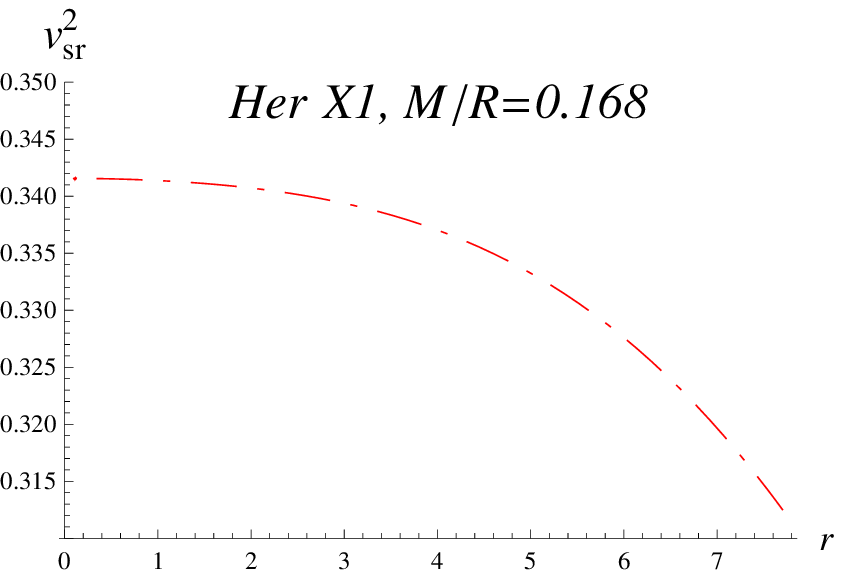, width=.34\linewidth,
height=1.4in}\epsfig{file=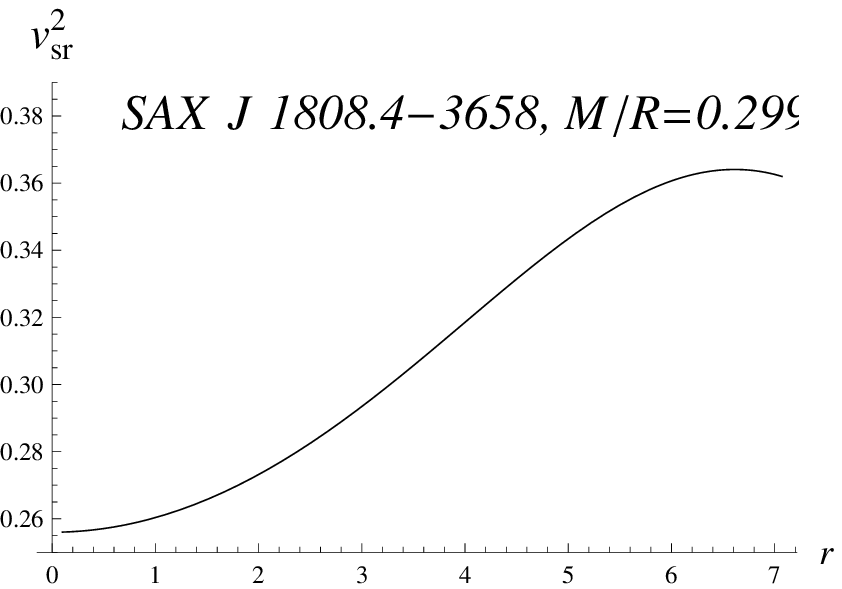, width=.36\linewidth,
height=1.4in}\epsfig{file=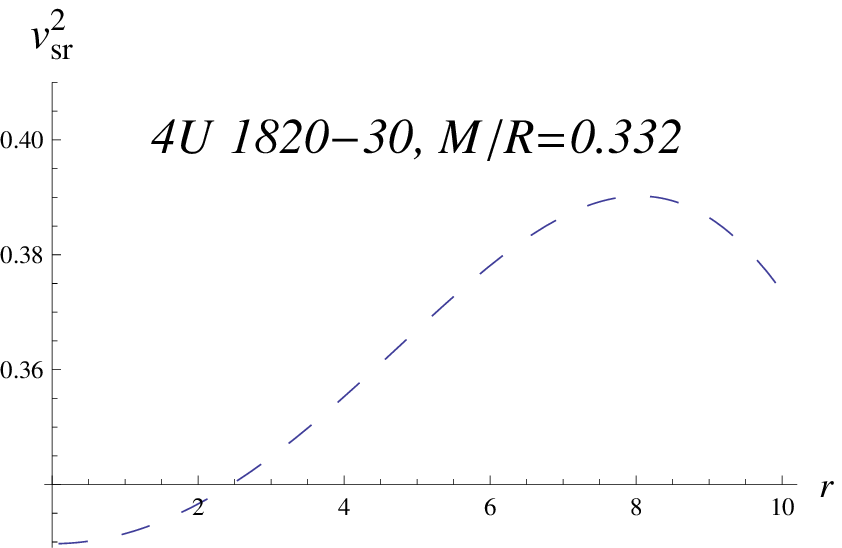, width=.34\linewidth,
height=1.4in}\caption{Variation of $v^2_{sr}$ for compact star
candidates.}
\end{figure}

The difference of $v^2_{sr}$ and $v^2_{st}$ can be obtained as
\begin{eqnarray}\nonumber
v^2_{st}-v^2_{sr}&=&\{e^{2Ar^2}(r^2-8\lambda)-4(-14+3Br^2+3B^3r^6+
2B^4r^8+12A^4r^8(2\\\nonumber&+&Br^2)-2A^3r^6(14+37Br^2+7B^2r^4)+A^2r^4(-22+37Br^2
+46\\\nonumber&\times&B^2r^4+4B^3r^6)-Ar^2(25-6Br^2+16B^2r^4+10B^3r^6+2B^4r^8))\\\nonumber&\times&\lambda-
e^{Ar^2}((A^2-AB+B^2)r^6+A(A-B)Br^8+48\lambda+Ar^4(1\\\nonumber&+&12A\lambda-12B\lambda)
+r^2(1+36A\lambda-12B\lambda))\}/\{e^{2Ar^2}(r^2-4\lambda)+4\\\nonumber&\times&(-5-3B^3r^6-B^4r^8
+12A^4r^8(2+Br^2)-A^3r^6(52+80Br^2\\\nonumber&+&11B^2r^4)+A^2r^4(-4+82Br^2+37B^2r^4-2B^3r^6)
+Ar^2(-7\\\nonumber&-&16B^2r^4+8B^3r^6+B^4r^8))\lambda+e^{Ar^2}(-Ar^4+2A^2r^6+24\lambda
\\\label{26}&+&r^2(-1+12A\lambda))\}.
\end{eqnarray}
The $v^2_{st}-v^2_{sr}$ of different strange stars is shown in
Figure \textbf{13}. Thus, our proposed model is stable.

\begin{figure}
\centering \epsfig{file=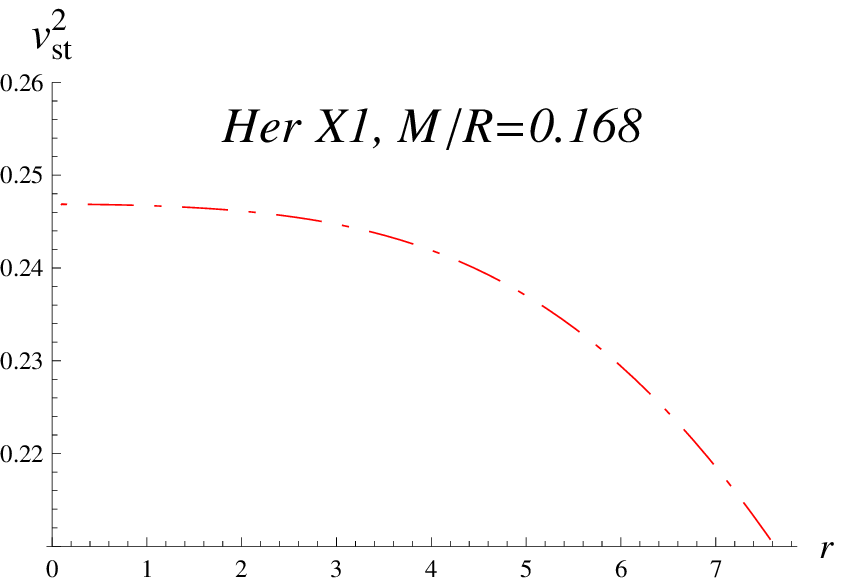, width=.34\linewidth,
height=1.4in}\epsfig{file=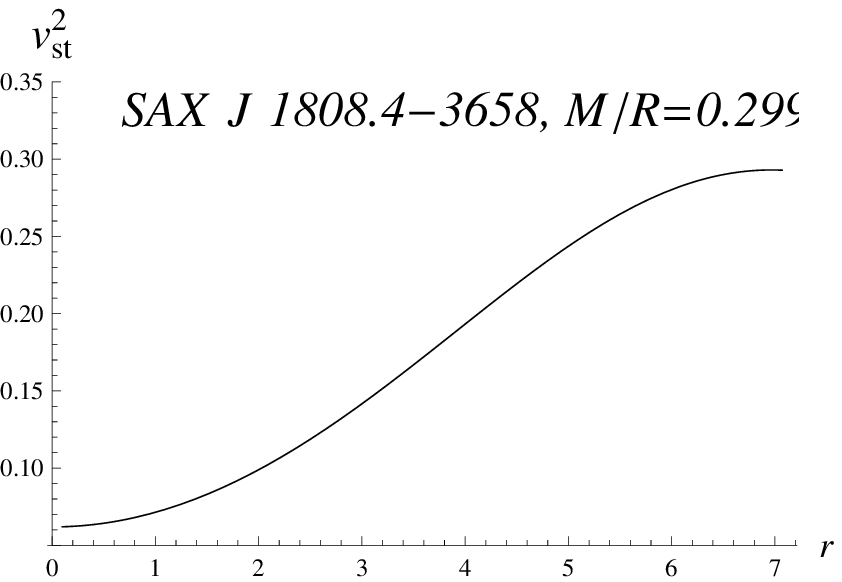, width=.36\linewidth,
height=1.4in}\epsfig{file=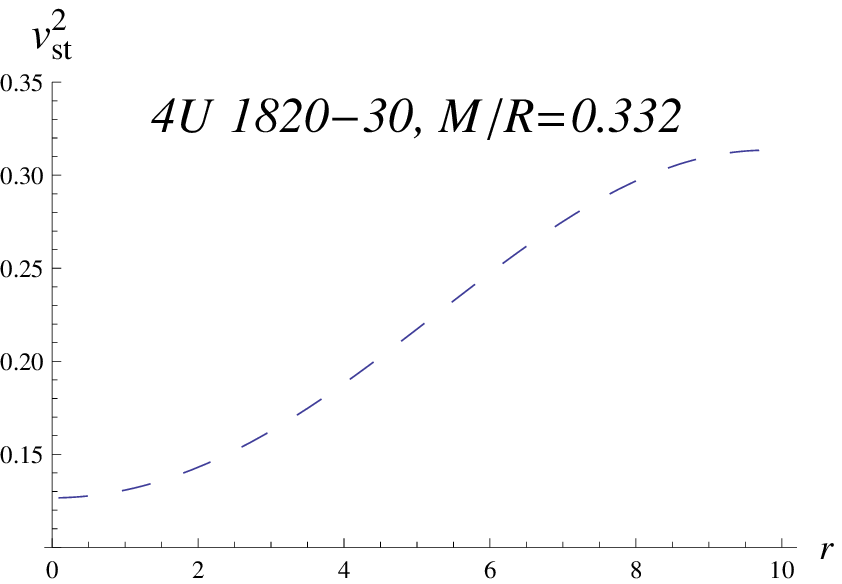, width=.34\linewidth,
height=1.4in}\caption{Variation of $v^2_{st}$  for compact star
candidates.}
\end{figure}
\begin{figure}
\centering \epsfig{file=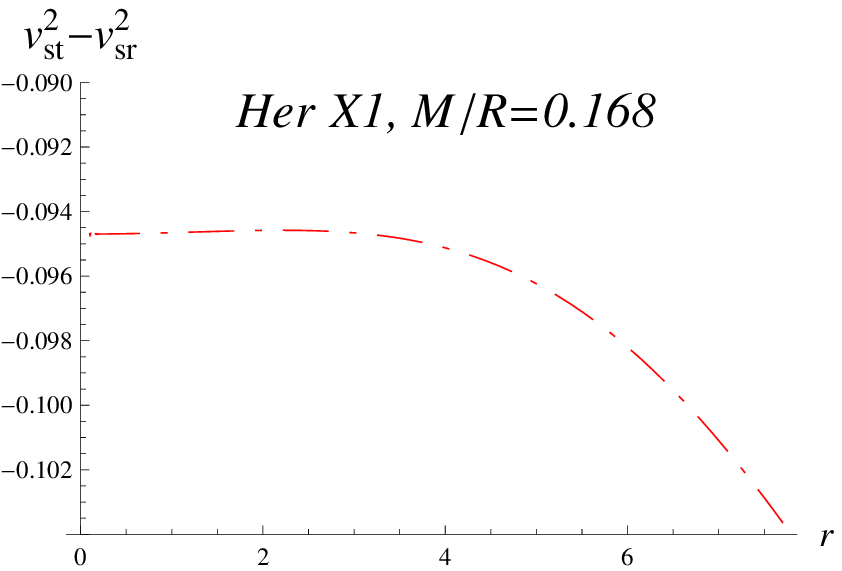, width=.34\linewidth,
height=1.4in}\epsfig{file=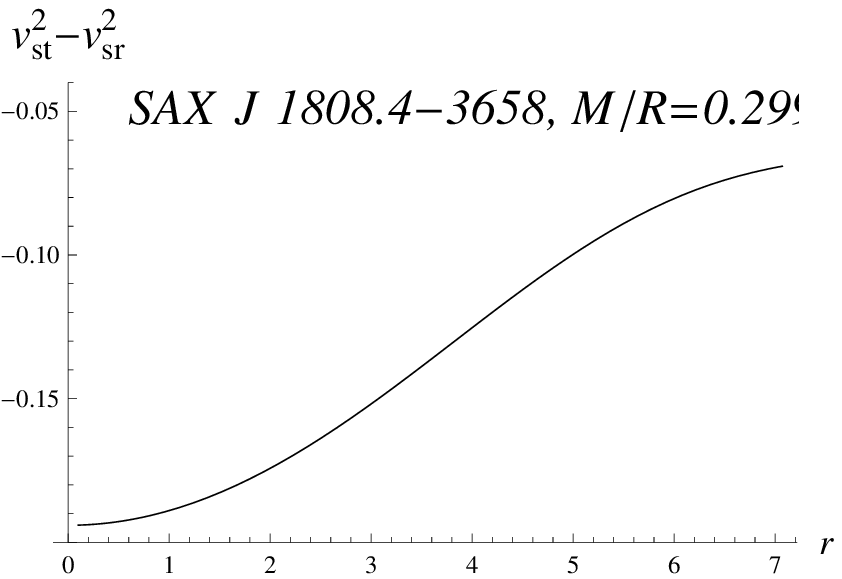, width=.36\linewidth,
height=1.4in}\epsfig{file=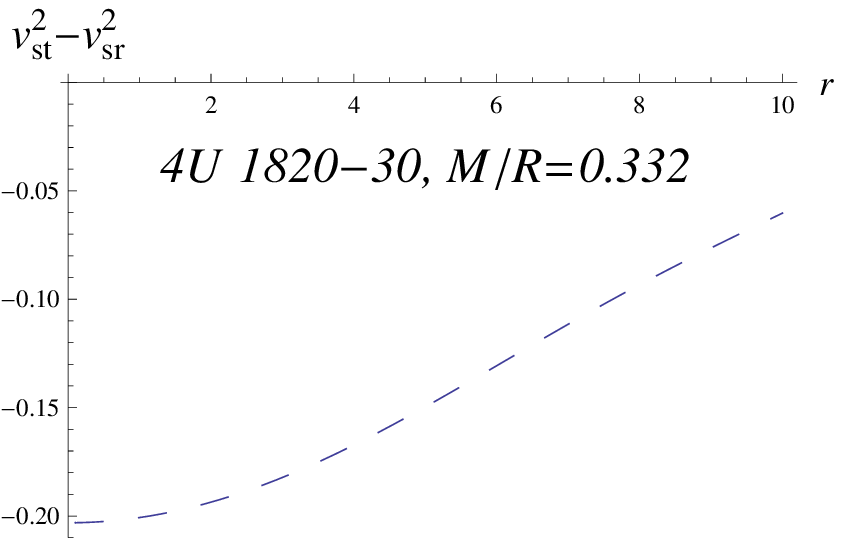, width=.34\linewidth,
height=1.4in}\caption{Variation of $v^2_{st}-v^2_{sr}$ for compact
star candidates.}
\end{figure}

\subsection{Surface Redshift}

The Mass-radius relation is
\begin{eqnarray}\nonumber
u&=&\frac{M(R)}{R}=\frac{{\pi}e^{-2AR^2}}{32A^2(AR^2)^{3/2}}\{-4\sqrt{A
R^2}(15R^2B^4\lambda+192A^5R^4(2+R^2B)\lambda+2AR^2\\\nonumber&\times&B^3(21+10R^2B)\lambda
-16A^4R^2(22+53R^2B+11R^4B^2)\lambda+A^2R^2B^2(99\\\nonumber&+&56R^2B+16R^4B^2)\lambda
-4A^3(-49R^4B^2\lambda+8R^6B^3\lambda+16(5-6e^{AR^2}\\\nonumber&+&e^{2AR^2})\lambda
+R^2(-8e^{AR^2}+8e^{2AR^2}-33B\lambda)))-1536A^4R^2e^{2AR^2}Erf(\sqrt{AR^2})\\\nonumber&\times&\sqrt{\pi}{\lambda}
+3R^2(224A^4+44A^3B+33A^2B^2+14AB^3+5B^4)e^{2AR^2}\sqrt{2\pi}{\lambda}
\\\label{27}&\times&Erf(\sqrt{2AR^2})\}.
\end{eqnarray}
The surface redshift ($Z_s$) is
\begin{eqnarray}\nonumber
{1+Z_s}&=&(1-2u)^{-1/2}=\{1-\frac{{\pi}e^{-2AR^2}}{16A^2(AR^2)^{3/2}}\{-4\sqrt{A
R^2}(15R^2B^4\lambda+192A^5R^4\\\nonumber&\times&(2+R^2B)\lambda+2AR^2B^3(21+10R^2B)\lambda
-16A^4R^2(22+53R^2B+11R^4\\\nonumber&\times&B^2)\lambda+A^2R^2B^2(99+56R^2B+16R^4B^2)\lambda
-4A^3(-49R^4B^2\lambda+8R^6B^3\\\nonumber&\times&\lambda+16(5-6e^{AR^2}e^{2AR^2})\lambda
+R^2(-8e^{AR^2}+8e^{2AR^2}-33B\lambda)))-1536A^4\\\nonumber&\times&R^2e^{2AR^2}Erf(\sqrt{AR^2})\sqrt{\pi}{\lambda}
+3R^2(224A^4+44A^3B+33A^2B^2+14AB^3\\\label{28}&+&5B^4)e^{2AR^2}\sqrt{2\pi}{\lambda}Erf(\sqrt{2AR^2})\}\}^{-1/2}.
\end{eqnarray}
Figure 13 shows the plot of redshift of compact star Her X-1 of
radius 7 km and the maximum redshift turns out to be $Z_s=0.845$.
\begin{figure}
\centering \epsfig{file=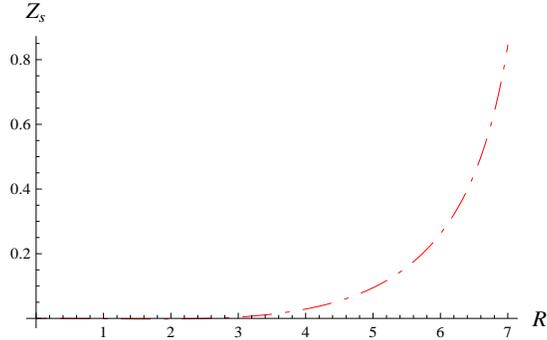, width=.52\linewidth,
height=1.8in}\caption{Surface redshift  of Her X-1.}
\end{figure}

\section{Conclusion}

The modified $f(R)$ theory of gravity providing the theoretical
explanation of accelerated expansion of universe, has attracted the
much attention of modern cosmologist. This theory has attained a
particular interest since the $f(R)$ modifications to general theory
of relativity appeared in a very natural way in the low-energy
effective actions of the quantum theory of gravity and the
quantization of underlying fields in curved spacetime. This theory
is also conformally related to GR with some exotic scalar field
\cite{33}.

This paper deals with the study of anisotropic compact stars whose
interior source is static. To complete the study, we have considered
that there may exists such compact stars that have anisotropy in
their interiors in the framework of $f(R)$ gravity. The interior
geometry of the compact stars has been handled by metric assumption
proposed by Krori and Barua \cite{18}. Then we perform the matching
of the interior metric with exterior Schwarzschild metric to
determine the constants of interior metric in terms of  of masses
and radii of the compact stars. The application of the masses and
radii of the compact stars yield the values of constants that
determine the nature of the stars. For these values of the
constants, we found that the energy conditions hold for the given
class of compact strange stars. By the physical interpretation of
the results, we conclude that the EOS parameters are given by
$0<\omega_i(r)<1$, ($i=r,t$). This indicates that fact that compact
stars are composed of ordinary matter and effect of $f(R)$ gravity
term. The matter components remains finite and positive every where
inside the stars and attain the maximum value at the center. Thus
our considered compact stars models are singularity free.

It is interesting to note that anisotropic force will be directed
outward when $P_t>P_r$ this implies that $\Delta>0$. We have found
that $\Delta>0$ for the different strange stars as shown in Figure
\textbf{8}. Hence, in this case repulsive force exists which allows
the construction of more massive stellar configuration in $f(R)$
gravity. The subliminal velocity of sound is less than 1,i.e, $0 <
v^2_{sr},$ $v^2_{st} < 1$ and $v^2_{sr} > v^2_{st}$. The variation
of $v^2_{st}-v^2_{sr}$ for different strange stars is shown in
Figure \textbf{13}, which satisfies the inequality
$|v^2_{st}-v^2_{sr}|\leq1$. Thus, in the presence of $f(R)$ term the
constructed compact stars models are stable. The range of surface
redshift $Z_s$ for the class of the particular star is
$0<Z_s\leq0.845$. The analysis of the compact stars in GR in the
absence of cosmological constant implies that redshift is
$Z_s\leq2$. Therefore, we conclude that in the present situation
redshift has been reduced to a certain value.

\vspace{0.25cm}

{\bf Conflict of Interests}

\vspace{0.25cm}

The authors declare that there is no conflict of interests
regarding the publication of this paper.


\vspace{.25cm}


\begin{thebibliography}{40}

\bibitem{1}Starobinsky, A.A.: Phys. Lett. B \textbf{91}(1980)99.

\bibitem{2}Nojiri, S. and Odintsov, S. D.: Phys. Rep. \textbf{505}(2011)59;
Bamba, K. Capozziello, S. Nojiri, S. and Odintsov, S.D.: Astrophys.
Space Sci. \textbf{345}(2012)155. Sharif, M. and Zubair, M.: JCAP
\textbf{11}(2013)042; Sharif, M. and Zubair, M.: J. High Energy
Phys. \textbf{12}(2013)079.

\bibitem{3}Capozziello, S., De Martino, I., Odintsov, S. D. and Stabile, A.:
Phys. Rev. D\textbf{83}(2011)064004.

\bibitem{4}Capozziello, S., De Laurentis, M., De Martino, I., Formisano, M. and
Odintsov, S. D.: Phys. Rev.D \textbf{85}(2012)044022.

\bibitem{5}Psaltis, D.: Living Reviews in Relativity
\textbf{11}(2008)9.

\bibitem{6}Briscese, F., Elizalde, E., Nojiri, S. and Odintsov, S. D.: Phys. Lett.
B \textbf{646}(2007)105; Nojiri, S. and Odintsov, S.D.: Phys. Lett.
B \textbf{676}(2009)94; Kobayashi, T. and Maeda, K.-i.: Phys. Rev. D
\textbf{78}(2008)064019.

\bibitem{7}Tsujikawa, S., Tamaki, T., and Tavakol, R.: JCAP
\textbf{05}(2009)020; Upadhye, A. and Hu, W.: Phys. Rev. D
\textbf{80}(2009)064002.

\bibitem{8}Arapoglu, S., Deliduman, C. and Eksi, K.Y.: JCAP
\textbf{07}(2011)020.

\bibitem{8a}Alavirad, H. and Weller, J.M.: Phys. Rev. D
\textbf{88}(2013)124034.

\bibitem{8b}Astashenok, A.V., Capozziello, S. and Odintsov, S.D.: Phys. Rev. D
\textbf{89}(2014)103509; JCAP \textbf{01}(2015)001.

\bibitem{8c}Astashenok, A.V., Capozziello, S. and Odintsov, S.D.: JCAP \textbf{01}(2015)001.

\bibitem{8d}Yazadjiev, S.S., Doneva, D.D., Kokkotas, K.D. and Staykov, K.V.: JCAP
\textbf{06}(2014)003.

\bibitem {9} Egeland,E. \textit{Compact Stars} (Trondheim, Norway, 2007).

\bibitem{15} MaK, M.K. and Harko, T.: Int. J. Mod. Phys. D \textbf{13}(2004)149.

\bibitem{16} Chaisi, M. and Maharaj S.D.: General Relativ. Gravit.
\textbf{37}(2005)1177.

\bibitem{17} Rahaman, F. et al.: Eur. Phys. J. C \textbf{72}(2012)2071.

\bibitem{18} Krori K.D. and Barua, J.: J. Phys. A.: Math. Gen.\textbf{8}(1975)508.

\bibitem{19} Lobo, F.S.N.: Class. Quant. Grav. \textbf{23}(2006)1525.

\bibitem{20} Alcock, C., Farhi, E. and Olinto, A.: Astrophys. J.
\textbf{310}(1986)261.

\bibitem{21} Haensel, P., Zdunik J.L. and Schaeffer, R. Astron. Astrophys.
\textbf{160}(1986)121.

\bibitem{28a} Hossein S.K.M.: et al, Int. J. Mod. Phys. D \textbf{21}(2012)1250088.

\bibitem{29a} Goswami et al.:Phys. Rev. D \textbf{90}(2014)084011.arXiv:1409.2371

\bibitem{29b} Cooney, A. et al.:Phys. Rev. D \textbf{83}(2010)064033.

\bibitem{29aa} Ganguly, A. et al.: Phys. Rev. D \textbf{89}(2014)064019

\bibitem{29b} Sharif, M. and Yousaf, Z.: MNRAS \textbf{440}(2014)3479.

\bibitem{29c} Sharif, M. and Yousaf, Z.: Astroparticle Phys. \textbf{56}(2014)19.

\bibitem{29d} Ifra, N. and Zubair, M.: Eur.Phys.J. \textbf{C75}(2015)62.

\bibitem{29e} Ifra, N. et al. : JCAP\textbf{1502}(2015)033.

\bibitem{1*}Li, X.-D. et al.: Phys. Rev. Lett. \textbf{83}(1999)3776.

\bibitem{2*}Zhang, W. et al.: ApJ \textbf{500}(1998)L171.

\bibitem{3*}Guver, T. et al.: ApJ \textbf{719}(2010)1807.

\bibitem{4*}Abubekerov, M.K. et al.: Astronomy Reports, \textbf{52}(2008)379.

\bibitem{29f}Lattimer, J.M. and Steiner, A.W.: Astrophys. J. \textbf{784}(2014)123 .

\bibitem{29g}Li, X.D., Bombaci, I., Dey, M., Dey J. and van den
Heuvel, E.P.J.: Phys. Rev. Lett. \textbf{83}(1999)3776.

\bibitem{30}Herrera L.: Phys. Lett. A, \textbf{165}(1992)206.

\bibitem{31}Chan, R. Herrera, L. and Santos, N.O.: MNRAS \textbf{265}(1993)533.

\bibitem{32}DiPrisco, A. Herrera, L. and Varela, V.: Gen. Relativ. Grav. \textbf{29}(1997)1239.

\bibitem{33} Barrow, J.D. and Cotsakis, S.: Phys. Lett. B\textbf{214}
(1988)515.

\end{thebibliography}
\end{document}